\AtBeginDocument{\RenewCommandCopy\qty\SI}
\documentclass[preprint,11pt,prb,amsmath,amssymb,aps,nobibnotes,nofootinbib,floatfix]{revtex4-2}

\usepackage{graphicx}
\usepackage{dcolumn}
\usepackage{bm}

\usepackage{physics}
\usepackage{tikz}
\usepackage{amsmath}
\usepackage{amssymb}
\usepackage{comment}
\usepackage{siunitx}
\usepackage{physics}
\usepackage{xfrac}
\usepackage{relsize}
\usepackage[none]{hyphenat}

\begin{document}

\title{Analytical Diagonalization of Fermi Gas-like Hamiltonians using the Sommerfeld-Watson Transformation}
\author{G. Diniz¹, F. D. Picoli¹, and M. P. Lenzarini¹}
\affiliation{¹Instituto de Física de São Carlos, Universidade de São Paulo, São Carlos, São Paulo, Brasil}%
\begin{abstract}
The Sommerfeld-Watson transformation is a powerful mathematical technique widely used in physics to simplify summations over discrete quantum numbers by converting them into contour integrals in the complex plane. This method has applications in scattering theory, high-energy physics, quantum field theory, and electrostatics. A lesser-known but significant use is in the analytical diagonalization of specific Hamiltonians in condensed matter physics, such as the Fermi gas Hamiltonian and the single-impurity Anderson model with vanishing Coulomb repulsion. These models are used to describe important phenomena like conductance in metals, x-ray photoemission, and aspects of the Kondo problem. In this work, we provide a comprehensive explanation of the Sommerfeld-Watson transformation and its application in diagonalization procedures for these models, using modern notation to enhance clarity for new students. The analytical results were validated against the numerical diagonalization, showing excellent agreement. Furthermore, we extend the presented method to a more generalized non-interacting single-impurity Anderson model with variable couplings and arbitrary band dispersion. The procedure presented here successfully achieved the analytical diagonalization of this more complex model, providing a unified solution that encompasses simpler cases. To our knowledge, this general solution has not been previously reported.
\end{abstract}
\maketitle
\newpage

\section{Introduction}

The Sommerfeld-Watson transformation is a mathematical technique used to convert a series to a contour integral in the complex plane, simplifying its evaluation \cite{sommerfeld1949partial}. Such series naturally arise in problems involving discrete quantum numbers, such as those encountered in scattering theory \cite{10.1121/1.402986}, where sums over poles or discrete quantum states are common. This technique is also applied in the analysis of helicity amplitudes in high-energy physics \cite{Goddard1971}, in the study of many-particle scattering theory \cite{PhysRev.134.B612}, and is frequently employed in quantum field theory, particularly in the analysis of spectral functions \cite{Frederiksen1975}. It can also be applied to electrostatic problems, where sums over discrete functions are comm \cite{10.1119/1.1975793,Saffet}. Undoubtedly, it is a highly useful method that expands the analytical toolkit of a theoretical physicist.

A lesser-known application of the Sommerfeld-Watson transformation is its use in the analytical diagonalization of specific Hamiltonians in condensed matter physics. For instance, the Fermi gas Hamiltonian, a fundamental and widely used model in many-body physics that describes phenomena such as electrical conductance in metals, the Anderson orthogonality catastrophe \cite{Anderson_1967},  x-ray emission and absorption \cite{Mahan,Nozieres,Nozieres71,oliveira1985}, and the x-ray photoemission problem \cite{Doniach_1970,ohtaka1990,libero1990,GDiniz,Picoli}. Additionally, Sommerfeld-Watson transformation can also be applied to diagonalize the single impurity Anderson model \cite{PhysRev.124.41}  in the non-interacting limit, i.e., when the Coulomb repulsion is turned off, the fixed-point Hamiltonians in the Renormalization Group theory, and to analyze fixed-point Hamiltonians in the Renormalization Group theory \cite{RevModPhys.47.773}, which play a fundamental role in the Fermi liquid description of the Kondo problem \cite{oliveira1994,PhysRevB.106.075129}.

In this work, our goal is to demonstrate that the Sommerfeld-Watson-based diagonalization procedure, previously applied to simple models, can be generalized and extended to more complex Hamiltonians. The examples presented here have already been studied in the references cited in the previous paragraph. However, these works often omit certain mathematical details or use outdated notations. Our first objective is to provide a comprehensive explanation of the Sommerfeld-Watson transformation and illustrate its role in the analytical diagonalization of the discussed models. We adopted a modern notation that is not always readily available in the existing literature, making this work a valuable reference for students learning this method. Furthermore, we compared the analytical results with numerical solutions obtained through brute-force diagonalization of these models across different parameter sets, revealing, as expected, an excellent agreement between the two approaches.

Finally, we consider a more general non-interacting single-impurity Anderson model, where we do not assume a specific dispersion for the band energies and allow the couplings in the Hamiltonian to vary, depending on the band levels they couple to. Despite the increased mathematical complexity, we demonstrate that the same diagonalization procedure used for simpler models remains effective in this more general case. Furthermore, we present a unified solution for this generalized non-interacting single-impurity Anderson model, from which the solutions for simpler cases can be derived. To the best of our knowledge, no previous work has derived the expression we obtained for this more general model.

This article is organized as follows: In Section II, we outline a simplified diagonalization procedure for analytically diagonalizing the Hamiltonians under consideration and describe the general form required for this approach. In Sections III and IV, we apply this procedure to two well-known examples: the Fermi gas and the non-interacting single-impurity Anderson model. In Section V, we compare the results from Sections III and IV with the numerical diagonalization of these models. Section VI introduces a general non-interacting single-impurity Anderson model and its analytical diagonalization. Finally, Section VII presents our conclusions.

\section{The Diagonalization procedure}

Here, we outline the key steps for analytically diagonalizing certain types of Hamiltonians using the Sommerfeld-Watson transformation. In the following sections, we will apply these general ideas explored here in specific examples. The diagonalization procedure works with any quadratic fermionic Hamiltonian that can be written \cite{Sakurai_Napolitano_2020,Mahan2010-xj} as
\begin{equation}\label{H_test}
\begin{aligned}
 \hat H = \sum_k \varepsilon_k \hat{c}_k^\dagger \hat{c}_k + \sum_{k,q}\left( M_{k} \bar M_{q} +  M_q^* \bar M_k^*  \right) \hat{c}_k^\dagger \hat{c}_{q} .
\end{aligned}
\end{equation}
$ \hat{c}_k^\dagger $ ($ \hat{c}_k $) are the creation (annihilation) operators for fermions with energy $ \varepsilon_k $, where $ k $ identifies the possible energy levels of the system. The product of the elements $ M_k $ and $ \bar{M}_q $ represent the couplings between the fermionic states, and ``$ ^* $" denotes the complex conjugation.

The condition of the coupling being written as $ M_k \bar{M}_q$ may seem overly restrictive at first glance, but many Hamiltonians of interest can be written in this form \cite{Anderson_1967,oliveira1994,G_Diniz,GDiniz}. Here, we will focus on one-dimensional Hamiltonians with $k$ scalar. However, this procedure can also be adapted to higher-dimensional Hamiltonians with couplings $M_{\vec{k}} \bar{M}_{\vec{q}}$. While the summation and the resulting expressions may become more complicated in higher dimensions, it remains feasible. The simpler higher-dimensional case is when the energy depends only on $ k = |\vec{k}| $ (spherical symmetry in $ d $ dimensions), converting the sums $ \sum_{\vec{k}} $ into sums over $ k $ \cite{Mahan2010-xj}.

It is important to note that similar steps can be adapted to diagonalize bosonic Hamiltonians; however, some sign differences may arise throughout the calculations presented here. In the following steps, we outline the main procedure for diagonalizing the fermionic Hamiltonian in equation \eqref{H_test}.

\subsection*{First step: The Ansatz}

A diagonal Hamiltonian in second quantization can be written as
\begin{equation}\label{DiagH}
   \hat H = \sum_{m} \epsilon_{m} \hat g_m^\dagger \hat g_m,
\end{equation}
where $\{\epsilon_m\}$ are its eigenvalues and $\{\hat g_m^\dagger \}$ one-particle \emph{eigenoperators} of the Hamiltonian, which creates a fermion with energy $\epsilon_m$. Diagonalizing the Hamiltonian \eqref{H_test} means finding its sets of $\{\epsilon_m\}$ and $\{\hat g_m^\dagger \}$, such that it can be compressed in the above form. Since the initial basis forms a complete vector space, the operator $ \hat{g}_m^\dagger $ can be written as a linear combination of the initial ones $ \{ \hat{c}_k^\dagger \} $ as  
\begin{eqnarray}\label{gm_test}
\hat g_m^\dagger= \sum_k u_{k,m}  \hat c_{k}^\dagger.
\end{eqnarray}
where $ u_{k,m} $ are the coefficients of the linear combination that have to be found in the diagonalization procedure. For simplification and without loss of generality, the coefficients $\{u_{k,m}\}$ are taken as real numbers.

\subsection*{Second step: Computing the commutators}


In this step, we compute the commutators $[H, \hat{c}_k^\dagger]$ and $[H, \hat{g}_m^\dagger]$. This computation is crucial as it enables us to derive a system of coupled equations, the solutions are the eigenvalues and eigenstates of the Hamiltonians. After some manipulations following the steps shown in Appendix \ref{Commutator,General}, we find
\begin{eqnarray}\label{[H,Ck]_test}
[\hat H,  \hat c_k^\dagger] = \left[ \varepsilon_k  \hat c_k^\dagger + \bar M_{k} \sum_q   M_{q} \hat c_q^\dagger  +  M_{k}^* \sum_q   \bar M_{q}^* \hat c_q^\dagger    \right],
\end{eqnarray}
and
\begin{eqnarray}\label{[H,gm]_test}
[\hat H, \hat g_m^\dagger] = \epsilon_{m} \hat g_m^\dagger.
\end{eqnarray}  
In the next step, we use these commutation relations to derive equations that determine the eigenvalues and eigenoperators of the Hamiltonian.

\subsection*{Third step: The eigenenergies and eigenoperators equations}

This step involves extensive manipulation of the above equations and more complex mathematical derivations, which we aim to avoid at this point. All details are explicitly presented in Appendix \ref{Eigenvalues_and_Eigenvectors}. Instead, we will summarize the main steps, which consist of manipulating equations \eqref{gm_test}, \eqref{[H,Ck]_test}, and \eqref{[H,gm]_test} 
\begin{eqnarray}\label{Conection*}
\left| 1 -  \sum_k \frac{ \bar M_k M_k}{\epsilon_m-\varepsilon_k} \right|^2  = \sum_k \frac{| M_k|^2}{\epsilon_m-\varepsilon_k} \sum_k \frac{| \bar M_k|^2}{\epsilon_m-\varepsilon_k}.
\end{eqnarray}

This equation represents the eigenvalue equation, and its solutions, $ \{\epsilon_m\} $, correspond to the eigenvalues of the Hamiltonian. A similar coupled equation involving these summations determines the coefficients $u_{k,m}$. This is precisely where the Sommerfeld-Watson transformation becomes highly useful, in the sense that it simplifies the sums, 
\begin{equation}\label{Motiv:1}
    S(\epsilon) = \sum_{k} \dfrac{|M_k|^2}{\epsilon-\varepsilon_k},~~~~ \bar S(\epsilon) = \sum_{k} \dfrac{|\bar M_k|^2}{\epsilon-\varepsilon_k}, ~~ \mathrm{and}~~  \tilde S(\epsilon) = \sum_{k} \dfrac{M_k \bar M_k}{\epsilon-\varepsilon_k}.
\end{equation}
After this, the equation $\left| 1- \tilde S(\epsilon)\right|^2 = S(\epsilon)\tilde S(\epsilon)  $ can be solved and the eigenvalues of $\hat H$ determined.

\subsection*{Las step: The Sommerfeld-Watson transformation}

\begin{figure}[htb!]
	\centering
	\includegraphics[scale=0.26]{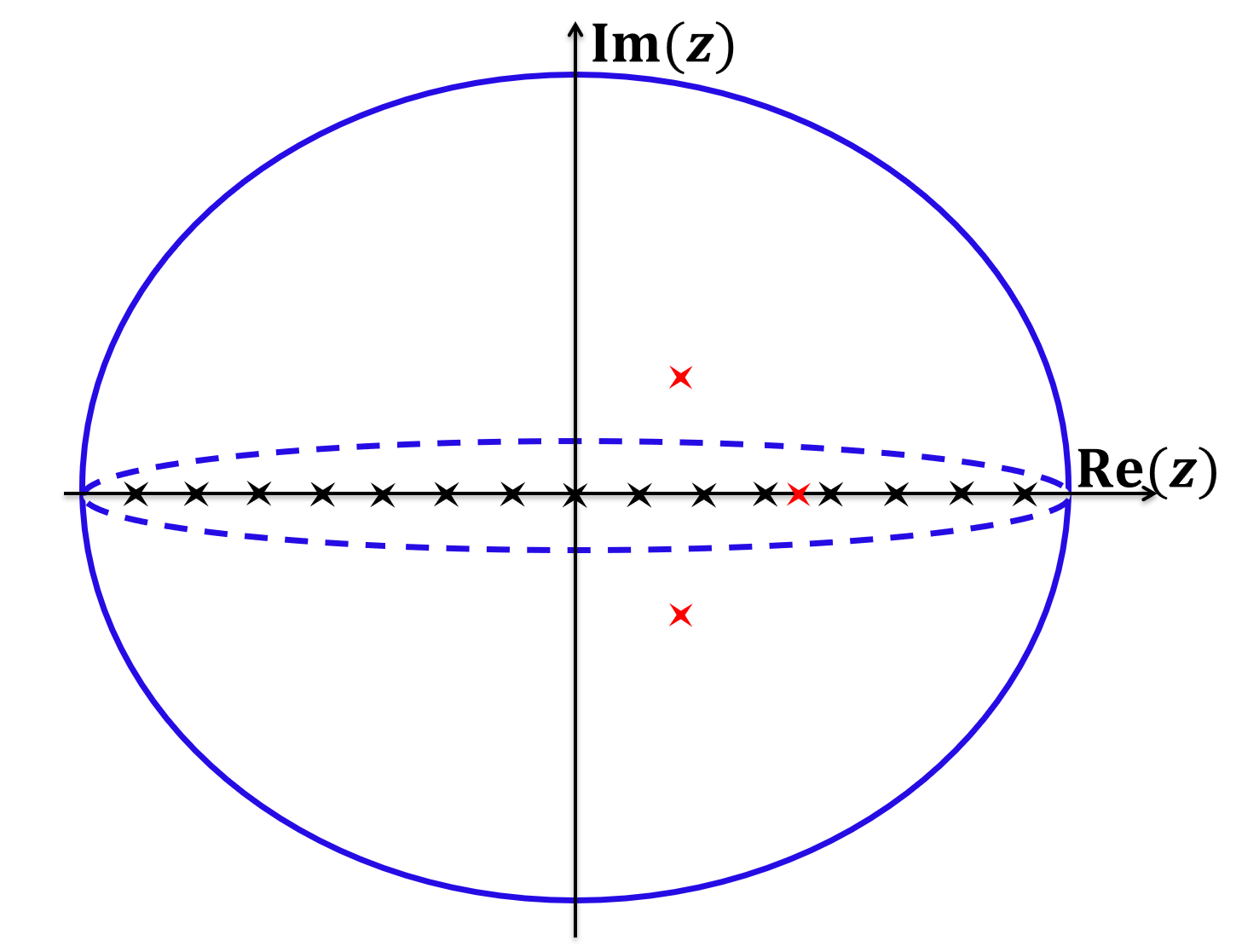}
        \includegraphics[scale=0.26]{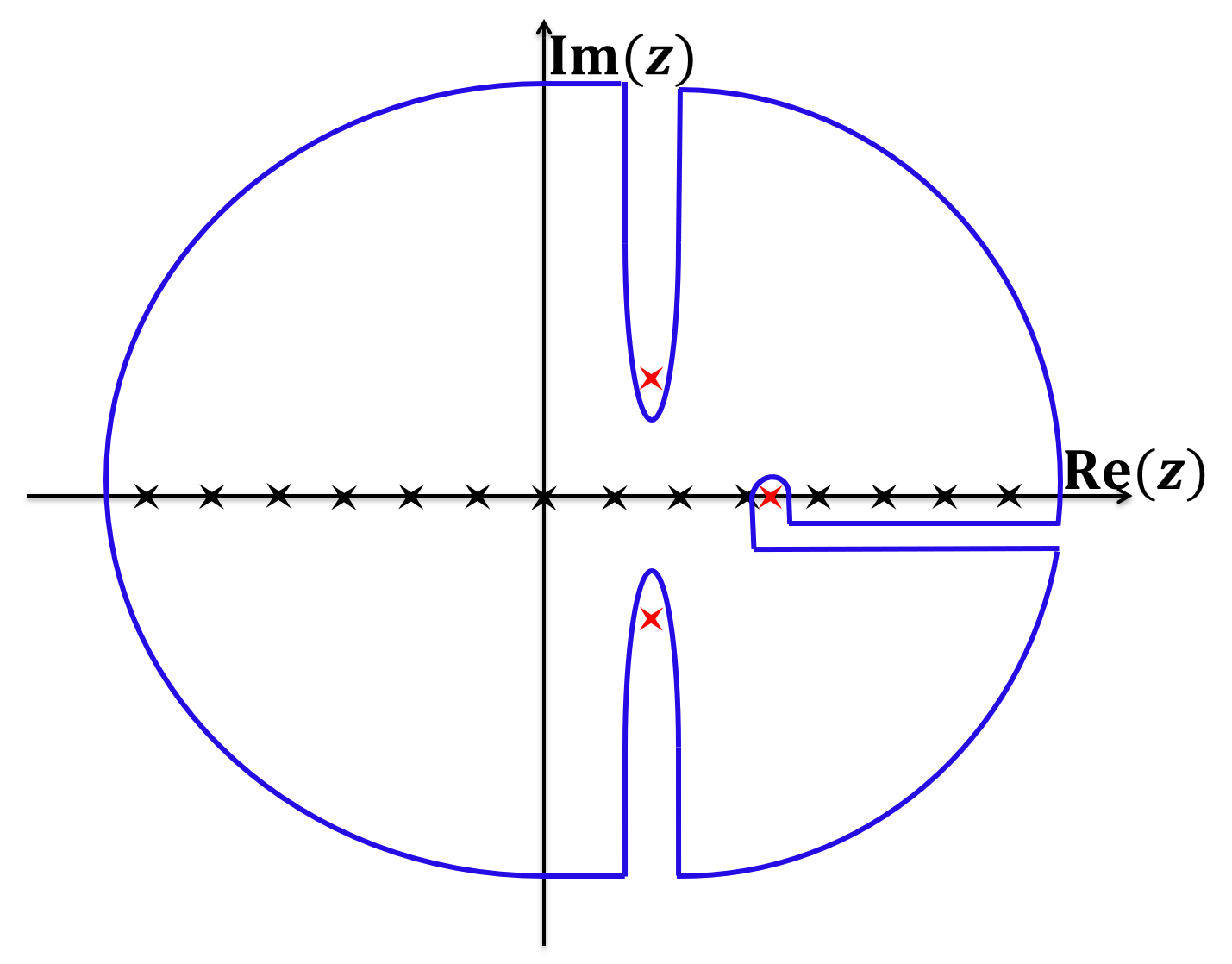}
        \vspace{-0.5cm}
	\caption{A few examples of appropriate contours for evaluating the Sommerfeld-Watson transformation are shown. The blue lines (solid and dashed) represent examples of suitable closed contours, while the black X symbols indicate the integer poles and the red X symbols denote the poles of the function $f(z)$. The latter poles do not necessarily need to be inside the contour.}
	\label{Paths}
\end{figure}

To complete the diagonalization, we address the sums in equation \eqref{Motiv:1}. Here, we focus on Fermi gas-like systems with a large number of particles, where these sums take the general form
\begin{align}\label{general_sum}
     S = \sum_{n=-\infty}^\infty f(n),
\end{align}
where $ f(n) $ is a function defined for discrete integers $ n $. The Sommerfeld-Watson transformation replaces this summation with the evaluation of a complex integral through the following steps. First, let us consider the complex function $F(z) = \pi f(z) \cot(\pi z)$, where $z \in \mathbb{C}$. $F(z)$ has simple poles in the set of integer values $\{n\}$ arising from the term $\cot(\pi z)$. The function can also have other poles coming from $f(z) $, which, for simplicity, we denote by $\{{z}_0\}$. Therefore, employing the residue theorem \cite{butkov1973mathematical}, we can define a contour integral in the complex plan: 
\begin{align}\label{SUM:AUX:I}
       \frac{1}{2 i}\oint_C {f(z)  \cot(\pi z)}\, dz = \sum_{n=-\infty}^\infty f(n) + \pi \sum_{\{z_0\} } \cot(\pi z_0) \mathrm{Res}( f(z),z=z_0),
\end{align}  
\noindent where the original sum we seek to evaluate in equation (\ref{general_sum}) appears among the residue terms. Here, we used the simple pole residue formula \cite{butkov1973mathematical}, that is, 
\begin{equation}
    \lim_{z \to n} f(z)\cot(\pi z) \cdot (z-n) = \lim_{z \to n} f(z)\cos(\pi z) \frac{(z-n)}{\sin(\pi z)} = \frac{1}{\pi} f(n).
\end{equation}
The contour $ C $ in equation \eqref{SUM:AUX:I} is an arbitrary closed complex contour that includes all integers, as shown in Fig. \ref{Paths}. In the above expression, $ \{ z_0 \} $ represents the set of all the poles of $f(z)$ within the contour $ C $. 

Moreover, by rearranging the equation \eqref{SUM:AUX:I}, we obtain
\begin{eqnarray}\label{SWT}
\sum_{n=-\infty}^\infty f(n)  = \frac{1}{2 i}\oint f(z)\cot(\pi z) dz - \pi \sum_{\{ z_0\}} \cot(\pi z_{0}) \mathrm{Res}(f(z),z=z_{0}).
\end{eqnarray}
To compute the right-hand side integral, we have to choose a convenient contour in the complex plane.

Small modifications to the transformation presented can be applied in other contexts. For example, by defining the function $ F(z) = \pi f(z) / \sin(\pi z) $, one can address alternating series. Additionally, the transformation can be adjusted to accommodate functions with integer poles, which requires evaluating higher-order residues at these poles and making the corresponding modifications to equation~\eqref{SWT}. In addition, the choice of the complex contour $C$ is arbitrary. A circular contour is especially appropriate for functions $f(z) $ with fast decay ($z^{-2}$ or faster), such as in the transformation used to solve the Bessel problem $S = \sum_{n \ge 1} n^{-2} $, where the contribution vanishes as the radius $ R \to \infty $.


Although there is no recipe for performing the Sommerfeld-Watson transformation, the procedure typically begins by defining the function $ F(z) = \pi f(z) \cot(\pi z) $ (or $ F(z) = \pi f(z) / \sin(\pi z) $ for alternating sums). This formulation isolates the poles of $ \sin(\pi z) $ from those of $ f(z) $, facilitating the extraction of the sum $ S $. The next steps involve selecting an appropriate closed complex contour, computing the residues of $ f(z) $, and deriving an expression for $ S $. Below, we present a few examples to illustrate the application of this method.

\section{Fermi Gas Under a Localized Scattering Potential}

In this section, we diagonalize the Fermi gas Hamiltonian in the presence of a localized scattering potential. Despite its simplicity, this Hamiltonian has played a crucial role in the development of new ideas and in explaining fascinating phenomena such as x-ray photoemission, leading to x-ray photoemission spectroscopy \cite{GRECZYNSKI2020100591}, and x-ray absorption and emission, leading to x-ray absorption spectroscopy \cite{Chantler2024}, among others.

Let us begin with a simpler version: the spinless Hamiltonian $\hat H$ in second quantization notation \cite{Sakurai_Napolitano_2020, Mahan2010-xj}, which describes a spinless Fermi gas interacting with a localized scattering potential of strength $W$ (an external charged particle or electric field, for example). It is given by
\begin{equation}\label{H_photo}
\begin{aligned}
\hat H = \sum_k \varepsilon_k \hat a_k^\dagger\hat a_k + \frac{W}{2L} \sum_{k,q}\hat a_k^\dagger\hat a_q.
\end{aligned}
\end{equation}
where $ \hat a_k^\dagger $ ($ \hat a_k $) are the creation (annihilation) operators for fermions in the momentum state $ k $, $2L$ is the number of states, and $\varepsilon_k $ is the energy dispersion relation. The first term corresponds to the effective kinetic energy of the Fermi gas, while the second term accounts for the scattering due to the localized potential of strength $W$. For simplicity, we will consider $\varepsilon_k = \Delta \cdot k$, a linear dispersion relation, where $\Delta$ is the difference between two consecutive energy levels, and $k$ is an integer such that $-L \le k \le+ L$. For a large number of electronic levels $\Delta \rightarrow 0$.

As mentioned in the previous section, the diagonal form of $\hat H$ is described by equation \eqref{DiagH}, where the operator $\hat g_m^\dagger$ can be written as a linear combination of the initial band operators $\{ \hat a_k^\dagger \}$ as
\begin{eqnarray}\label{gm}
\hat g_m^\dagger= \sum_k u_{k,m}  \hat a_k^\dagger.
\end{eqnarray}

Our task now is to determine the eigenvalues $\{\epsilon_m\}$ and the ``eigenvectors" coefficients $\{u_{k,m}\}$. Then, the second step of the procedure continues by computing $[\hat H, \hat a_k^\dagger]$ and $[\hat H,\hat g_m^\dagger]$. After some manipulations using appendix \ref{Commutator,General}, we arrive at the following expressions:
\begin{eqnarray}\label{[H,Ck]}
[\hat H,  \hat a_k^\dagger] = \left[ \varepsilon_k  \hat a_k^\dagger + \frac{W}{2L} \sum_q  \hat a_q^\dagger    \right]
\end{eqnarray}
and
\begin{eqnarray}\label{[H,gm]}
[\hat H,\hat g_m^\dagger] = \epsilon_{m}\hat g_m^\dagger.
\end{eqnarray}

Before proceeding, it is important to note that the scattering potential $W$ in the Hamiltonian $ \hat{H} $ shifts each initial eigenvalue $ \varepsilon_m $ by $ \left(-\dfrac{\delta_m }{\pi} \right) \Delta $, where $ \delta_m $ is the \emph{phase shift} and $ \Delta $ is the difference between two consecutive energy levels near the Fermi energy. In principle, $ \delta_m $ can take any value that adjusts the initial eigenvalue to match the eigenvalue $ \epsilon_m $ in the presence of the scattering potential. In this situation, we can write
\begin{eqnarray}\label{eigenvalues}
\epsilon_m = \varepsilon_m - \frac{\delta_m}{\pi} \Delta.
\end{eqnarray}

Using the equality defined in equation \eqref{[H,gm]} by substituting $\hat g_m^\dagger$ with the linear combination defined in \eqref{gm} and after some algebraic manipulations, we find
\begin{eqnarray}
\sum_k \left[\varepsilon_k u_{k,m} +\frac{W}{2L} \sum_q u_{q,m} \right]  \hat a_k^\dagger = \sum_k \epsilon_m u_{k,m} \hat a_k^\dagger.
\end{eqnarray}

As the operators $ \hat a_k^\dagger$ are linearly independent, each term of the $k$-sum above must satisfy the equation independently,
\begin{eqnarray}\label{E__I}
\left( \epsilon_m-\varepsilon_k \right) u_{k,m} - \frac{W}{2L} \sum_q u_{q,m} = 0.
\end{eqnarray}
Summing over the $k$-states, we can write
\begin{eqnarray}\label{E__II}
\sum_k u_{k,m} = W \left( \frac{1}{2L} \sum_k \frac{1}{\left(\epsilon_m-\varepsilon_k \right)} \right) \sum_q u_{q,m}.
\end{eqnarray}
If $\sum_k  u_{k,m}  \ne 0$, we obtain
\begin{eqnarray}\label{Ktau}
1 = W \left( \frac{1}{2L} \sum_k \frac{1}{\left(\epsilon_m-\varepsilon_k \right)} \right).
\end{eqnarray}

We wish now to analyze the sum on the right-hand side of Eq. (\ref{Ktau}). It is convenient to define a variable $S_m$ such that:
\begin{eqnarray}\label{S_m1}
S_m &= \dfrac{1}{2L} \mathlarger{\sum_q} \dfrac{1}{\epsilon_m-\varepsilon_q}.
\end{eqnarray}
Now, using equation \eqref{eigenvalues}, the summation $S_m$ given by the equation (\ref{S_m1}) can be rewrite as
\begin{eqnarray}
S_m &= \dfrac{1}{2L} \mathlarger{\sum_q} \dfrac{1}{\varepsilon_m-\varepsilon_q - \frac{\delta_m}{\pi} \Delta}.
\end{eqnarray}

Considering the \emph{bandwidth energy} as $D = \Delta \cdot L$, the metallic band \emph{density of states} is 
\begin{eqnarray}\label{Flatband_dos}
\rho \equiv \dfrac{1}{2D} = \dfrac{1}{(2L \Delta)},
\end{eqnarray}
and, after some manipulation, $S_m$ becomes:
\begin{eqnarray}\label{AUX_SWSUM_I}
S_m =  - \rho \sum_q \dfrac{1}{q - m + \frac{\delta_m}{\pi}}.
\end{eqnarray}

To solve the sum in equation \eqref{AUX_SWSUM_I}, we can use the Sommerfeld-Watson transformation \cite{sommerfeld1949partial}. As we already discussed,  considering a sum of the form $\sum_n f(n)$, where $f(z)$ is a function with noninteger poles, we can define a function $F(z) = \pi f(z) \cot (\pi z)$. The function $F(z)$ has simple integer poles $\{n\}$ coming from the term $\cot (\pi z)$ and noninteger poles $\{z_{n_i}\}$ coming from the term $f(z)$. We can apply this transformation to find the value of $S_m$, since we can write $S_m = \rho \sum_n f(n)$ with $f(z)=-(z-z_0)^{-1}$, a function with a single simple pole at $z_0= m-{\delta_m}/{\pi}$. After the Sommerfeld-Watson transformation, the equation (\ref{AUX_SWSUM_I}) results in
\begin{eqnarray}\label{ComplexIntegral}
 S_m = -\frac{1}{2 i}\oint\frac{dz~\rho~\mathrm{cot} (\pi z)}{z-\left(m-\frac{\delta_m}{\pi}\right)}+ \pi \rho \mathrm{cot}  (\pi m - \delta_m). ~~~~
\end{eqnarray}

\begin{figure}[htb!]
	\centering
	\includegraphics[scale=0.39]{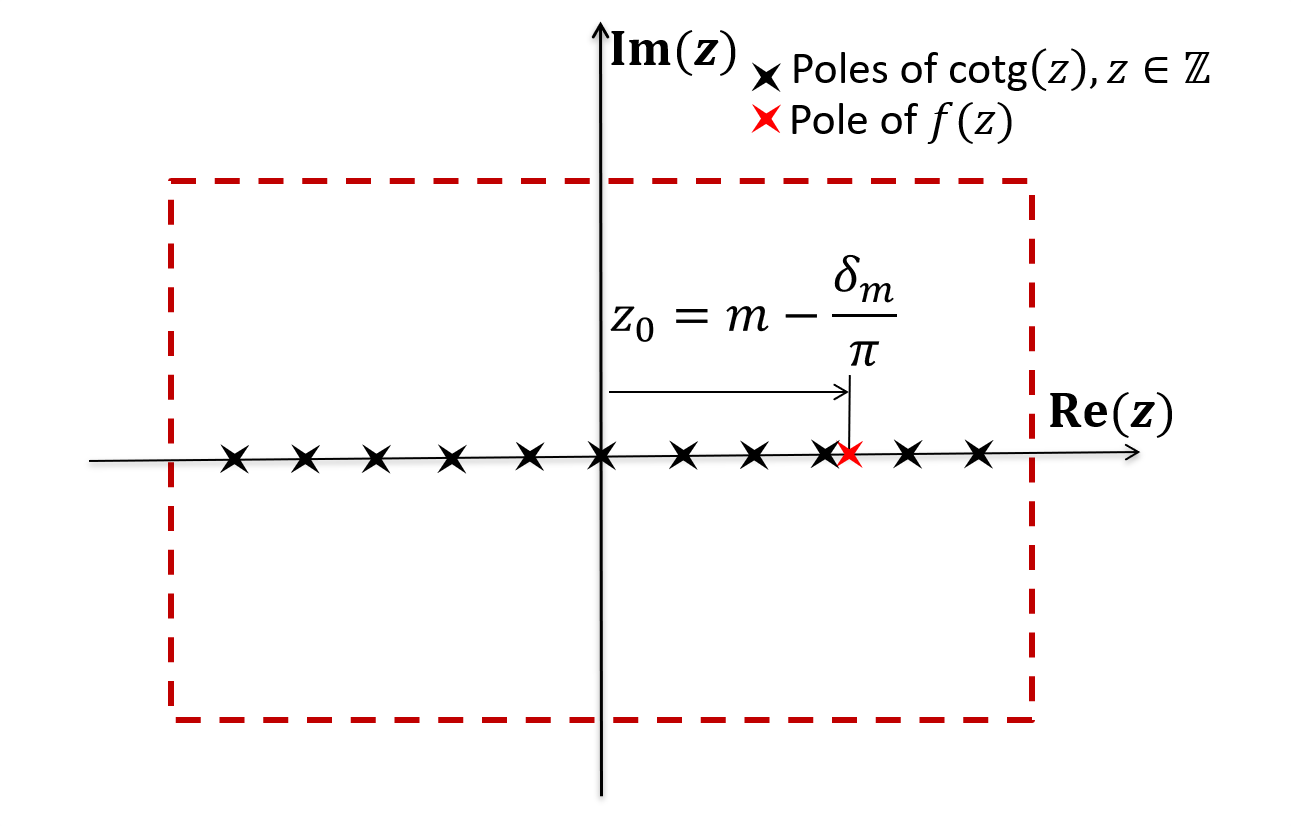}
        \includegraphics[scale=0.39]{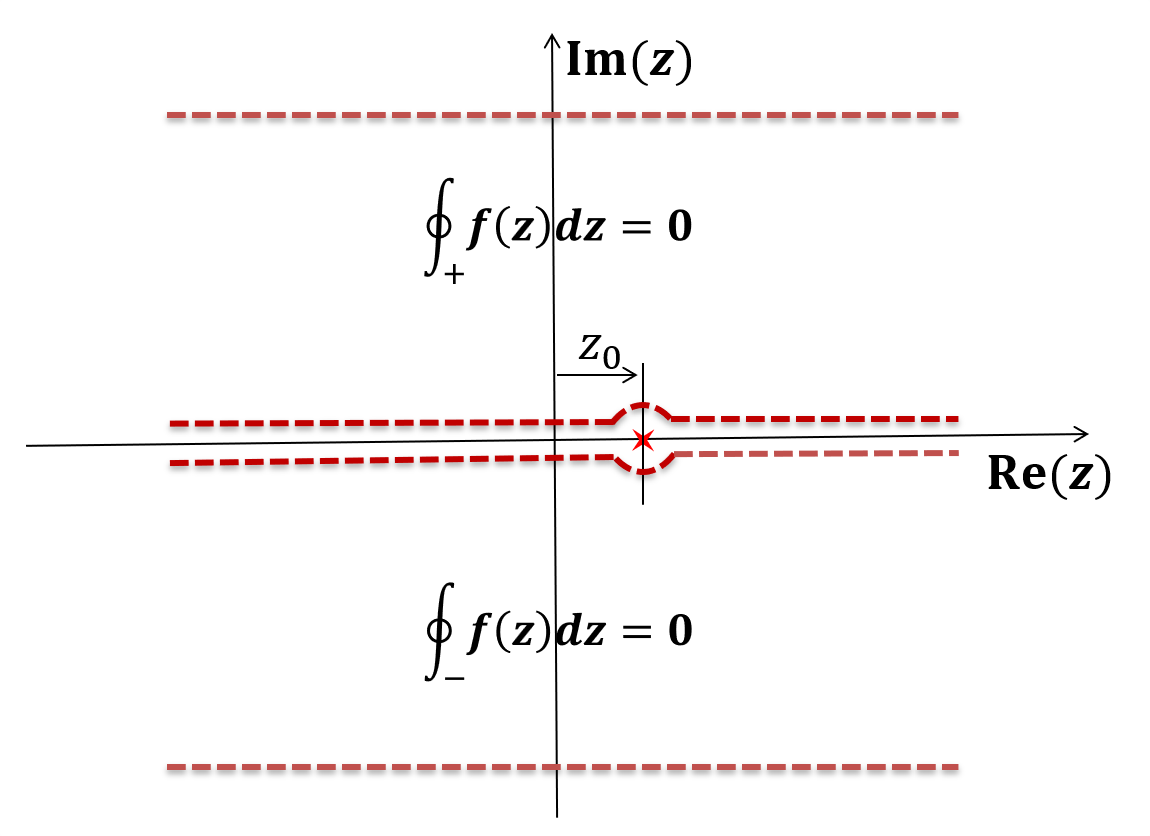}
        \vspace{-0.5cm}
	\caption{ (Left panel) Integration path of \eqref{ComplexIntegral}. (Right panel) In the upper and bottom boundaries of the integration path $\lim_{(\mathrm{Im} z \rightarrow \pm \infty)}\cot(\mathrm{Re}(z) +i\mathrm{Im} z) = \pm -i$, which results in the expression in equation \eqref{soma}. }
	\label{Sommerfeld-Watson}
\end{figure}

The last term can be simplified to $\cot (\pi m - \delta_m) = -\cot (\delta_m)$. Now we only need to find the integral. For this purpose, it is necessary to define a closed contour in the domain of the function $F(z)=f(z) \mathrm{cot} (z)$. An appropriate complex contour for this problem is shown in Fig. \ref{Sommerfeld-Watson} (left panel), once $\lim_{(\mathrm{Im} z \rightarrow \pm \infty)}\cot(\mathrm{Re}(z) +i \ \mathrm{Im} (z)) = \mp i$. Taking the limit where the box size shown in Fig. \ref{Sommerfeld-Watson} tends to infinity, we can write
\begin{eqnarray}\label{Complex_integral_II}
  \frac{1}{2 i}\oint dz\frac{\mathrm{cot} (\pi z)}{z-\left(m-\frac{\delta_m}{\pi}\right)} = -\frac{1}{2}\lim_{y \rightarrow + \infty }  \int_{+\infty}^{-\infty} dx \frac{1}{x+iy-z_0} +\frac{1}{2}\lim_{y \rightarrow - \infty }  \int_{-\infty}^{+\infty} dx \frac{1}{x+iy-z_0}.
\end{eqnarray}

Using an auxiliary path, shown in the right panel of Fig. \ref{Sommerfeld-Watson}, for a function $f(z)$ with poles only in the real axis, it is true that $\frac{1}{2 i}\oint_\pm f(z) dz  = 0 $. Moreover, it is possible to show that
\begin{eqnarray}\label{Complex_integral_III}
\lim_{y \rightarrow \pm \infty } \int_{+\infty}^{-\infty} dx \frac{1}{x+iy-z_0} = - \lim_{L \rightarrow \infty }\lim_{y \rightarrow  0^{\pm} }  \int_{-L}^{+L} dx \frac{1}{x+iy-z_0}.
\end{eqnarray}

Therefore, with the results in equations \eqref{Complex_integral_II} and \eqref{Complex_integral_III}, we found that the contribution from the closed path in Fig. \ref{Sommerfeld-Watson} to $S_m$ is 
\begin{equation}
    \mathcal{P}\mathlarger{\int}_{-L\Delta}^{+L \Delta} d\epsilon \dfrac{\rho }{\varepsilon_m - \epsilon},
\end{equation}
where $\mathcal{P}$ represents the Cauchy principal value of the integral. Then, the expression in equation \eqref{ComplexIntegral} becomes
\begin{eqnarray}\label{soma}
S_m =  - \pi\rho \mathrm{cot}(\delta_m) + \rho \mathcal{P} \int_{-L \Delta}^{+L \Delta} \dfrac{d\epsilon}{\varepsilon_m - \epsilon}. 
\end{eqnarray}

Once we have determined the value of $S_m$, we can return to the diagonalization procedure. From equation (\ref{Ktau}), we have $ W S_m = 1$. Then, by using the equation \eqref{soma}, we have
\begin{eqnarray}\label{phase_shift_equation}
\cot(\delta_m) &=& \frac{1}{\pi}\mathcal{P}\int_{-D}^{+D}\frac{1}{\varepsilon_m-\epsilon} d\epsilon  - \dfrac{1}{\pi \rho W} + \mathcal{O}(\Delta) \nonumber \\
~ \nonumber \\
&=& - \dfrac{1}{\pi \rho W} + \frac{1}{\pi}\ln \left( \frac{D + \varepsilon_m }{D - \varepsilon_m } \right).   
\end{eqnarray}
From this expression, it is possible to find the value of $\delta_m \equiv \delta_m(\varepsilon_m, W)$ for each $\varepsilon_m$ and fixed $W$, and consequently determine the energy eigenvalues using equation \eqref{eigenvalues}.


Our goal here is to determine the new energy levels in the presence of the scattering potential. To achieve this, we need to compute the phase shift $\delta_m$. From equation \eqref{phase_shift_equation}, the phase shift is given by
\begin{eqnarray}\label{phase_shift}
&\tan(\delta_m) = - {\pi \rho W} \left(1 - \rho W \ln \left( \dfrac{D + \varepsilon_m }{D - \varepsilon_m } \right) \right)^{-1},
\end{eqnarray}
or, for the energy levels close to the Fermi energy, the phase shift can be approximated as 
\begin{eqnarray}\label{phase_shift_ap}
&\tan(\delta_m) \approx \tan (\delta_0) \equiv - {\pi \rho W},
\end{eqnarray}
a constant that depends only on $W$, for $|m| \ll L$.

To complete the diagonalization of the Hamiltonian, we have to find the coefficients $u_{k,m}$ and, consequently, the eigenoperators $\{ g^\dagger_m\}$. We start by isolating the coefficient $u_{k,m}$ in equation \eqref{E__I} and squaring both sides, and then applying the sum over $k$. This gives
\begin{eqnarray}
\sum_k |u_{k,m}|^2=\frac{W^2}{2L} \frac{1}{2L} \sum_k \frac{1}{\left( \epsilon_m-\varepsilon_k \right) ^2} \left( \sum_q u_{q,m} \right)^2. \nonumber
\end{eqnarray}
The normalization of the single particle state guarantees that $\sum_k |u_{k,m}|^2 = 1$. Note that from equations \eqref{AUX_SWSUM_I} and \eqref{soma}, it is straightforward to show that 
\begin{eqnarray}\label{Derivative_sum}
 -\dv{S_m}{\epsilon_m}  = \frac{1}{2L} \sum_k \frac{1}{\left( \epsilon_m-\varepsilon_k \right) ^2} = \frac{1}{2L \Delta^2 }\left|\frac{\pi}{\sin \delta_m}\right|^2.
\end{eqnarray}
This leads to
\begin{eqnarray}
1={ \rho^2 W^2} \left(\left|\frac{\pi}{\sin \delta_m}\right|^2\right) \left( \sum_q u_{q,m} \right)^2, \nonumber \end{eqnarray}
or after taking the square root 
\begin{eqnarray}\label{sum_coef}
 \sum_q u_{q,m} = \pm \frac{\sin \delta_m}{\pi \rho W }.\end{eqnarray}

Finally, substituting the equation \eqref{sum_coef} into the equation \eqref{E__I} we obtain the coefficients $u_{k,m}$ as
\begin{eqnarray}\label{coeficients}
 u_{k,m} = -\frac{\Delta}{\left( \epsilon_m-\varepsilon_k \right)} \frac{\sin \delta_m}{\pi}, 
\end{eqnarray}
and the energy eigenvalues $\epsilon_m$ from equations \eqref{eigenvalues} and \eqref{phase_shift}, completing the diagonalization procedure. The minus sign in equation \eqref{sum_coef} is chosen to ensure that $\lim_{\delta \rightarrow 0} u_{k,m} = \delta_{k,m}^K$, where $\delta_{k,m}^K$ represents the Kronecker delta.

\section{The non-interacting Single Impurity Anderson Model}

Another important model in condensed matter physics is the single impurity Anderson model (SIAM) \cite{PhysRev.124.41}, which is extremely useful for understanding the behavior of quantum impurities coupled to a metallic conduction band. In the second quantization, this model can be described by the Hamiltonian
\begin{eqnarray}\label{AM}
 \hat H_{A} \equiv \left[ \varepsilon_d \left(\hat n_{d\uparrow} +\hat n_{d\downarrow} \right) + U \hat n_{d\uparrow} \hat n_{d\downarrow} \right] 
+ \hat H_B + \hat H_{\mathrm{hyb.}}.
\end{eqnarray}
The first term in brackets on the right-hand side is the contribution of impurity, where $\hat n_{d\sigma} = \hat d_\sigma^\dagger \hat d_\sigma$, and $\hat d_{\sigma}^\dagger (\hat d_{\sigma})$  creates (annihilates) one electron in the impurity level with energy $\varepsilon_d$ and spin $\sigma = \{\downarrow, \uparrow\}$. The Coulomb repulsion term penalizes double occupancy. Consequently, the impurity can have one of three energies, depending on the impurity occupancy (single, empty, or double). $\hat H_B$ represents the contribution of the conduction band, which can be considered a Fermi gas as shown in Section III. The hybridization term $\hat H_{\mathrm{hyb.}}$ is the coupling between the impurity and the metal, allowing electron transfer between the impurity and the metal.

In real space, it is a good approximation to consider that the impurity couples only to the neighboring metallic atoms. In momentum space, the coupling is constant, and the hybridization term becomes:
\begin{eqnarray}
 \hat H_{\mathrm{hyb.}} \equiv \frac{V}{\sqrt{2L}} \sum_{k\sigma}\left(\hat a_{k\sigma}^\dagger \hat d_\sigma + \hat d^\dagger_\sigma \hat a_{k\sigma} \right).
\end{eqnarray}

Despite the seemingly simple form of the SIAM defined in equation \eqref{AM}, the Coulomb interaction $ U $ makes the model non-quadratic, which cannot be diagonalized easily. To address the interacting case ($ U \neq 0 $), more sophisticated approaches, such as the Bethe Ansatz \cite{Schlottmann1983} or numerical methods like the Numerical Renormalization Group (NRG) \cite{libero1990,PhysRevB.101.125115}, the Real Space Numerical Renormalization Group (eNRG) \cite{Picoli,PhysRevB.106.075129}, and the Density Matrix Renormalization Group (DMRG) \cite{RevModPhys.47.773,RevModPhys.80.395,PhysRevLett.69.2863,PhysRevLett.93.076401}, are required. However, these methods lie beyond the scope of this paper. Here, we will focus on analytically diagonalizing the non-interacting SIAM (NI-SIAM), defined by setting $U=0$. For this case, each spin component $\sigma$ is completely independent of each other, and we can consider spinless fermions.

\subsection{Analytical diagonalization of the NI-SIAM}

It is beneficial to introduce a small modification to the Hamiltonian by incorporating a localized scattering potential term as used in the Fermi gas problem above, $\frac{W}{2L}\sum_{k,q} \hat a_k^\dagger \hat a_q$. This adjustment allows the Hamiltonian to easily recover the pure non-interacting SIAM case by setting $W = 0$, while also enabling its application to more general scenarios, such as the low-temperature Fermi liquid description of the Kondo problem \cite{oliveira1994,oliveira1985,PhysRevB.106.075129}. In this case, the new spinless Hamiltonian (also called as Newns-Anderson model \cite{PhysRev.178.1123}) can be written as:
\begin{eqnarray}
\hat H =\varepsilon_d \hat d^\dagger \hat d +\sum_{k}\varepsilon_{k} \hat a^\dagger_{k}\hat a_k +\frac{V}{\sqrt{2L}}\sum_{k} \left(\hat d^\dagger \hat a_k+{\mathrm H.c.} \right) +\frac{W}{2L}\sum_{k,q} \hat a_k^\dagger \hat a_q.
\end{eqnarray}
Similarly, as considered in the Fermi gas problem, we can diagonalize this Hamiltonian if it is expressed in the form $\hat H = \sum_{m} \epsilon_{m} \hat g_m^\dagger \hat g_m$, where the operator $\hat g_m^\dagger$ is a linear combination of the operators ${\hat a_k^\dagger}$ and $\hat{d}^\dagger$:
\begin{eqnarray}\label{gm2}
\hat g_m^\dagger = \sum_k u_{k,m} \hat a_k^\dagger + u_{d,m} \hat d^\dagger.
\end{eqnarray}


Now, following the diagonalization procedure described in Section II, we need to compute the commutators of $\hat H$ with $\hat g_m^\dagger$, $\hat a_k^\dagger$ and $\hat d^\dagger$. Using the result from Appendix \ref{Commutator,General}, after some manipulations, we arrive at the following expressions:
\begin{eqnarray}\label{[H2,Ck]}
[\hat H,\hat a_k^\dagger] = \left[ \frac{V}{\sqrt{2L}} \hat d^\dagger + \varepsilon_k \hat a_k^\dagger + \frac{W}{2L} \sum_q \hat a_q^\dagger    \right],
\end{eqnarray}
\begin{eqnarray}\label{[H2,Cd]}
[\hat H, \hat d^\dagger] = \left[ \varepsilon_d \hat d^\dagger + \frac{V}{\sqrt {2L}} \sum_q \hat a_q^\dagger  \right],
\end{eqnarray}
\begin{eqnarray}\label{[H2,gm2]}
[\hat H,\hat g_m^\dagger] = \epsilon_{m}\hat g_m^\dagger.
\end{eqnarray}


Using the equation \eqref{[H2,gm2]} and substituting the operator $\hat{g}_m^\dagger$ expanded as the linear combination defined in \eqref{gm2}, we arrive at
\begin{eqnarray}
\sum_k \left[ \varepsilon_k u_{k,m} +\frac{W}{2L} \sum_q u_{q,m} + \frac{V}{\sqrt{2L}} u_{d,m}  \right] \hat a_k^\dagger + \left[ \varepsilon_d u_{d,m} + \frac{V}{\sqrt {2L}} \sum_q u_{q,m} \right] \hat{d}^\dagger = \nonumber \\ \sum_k \epsilon_m u_{k,m}\hat a_k^\dagger +\epsilon_m u_{d,m}\hat{d}^\dagger.
\end{eqnarray}
To derive the above expression, we used equations \eqref{[H2,Ck]} and \eqref{[H2,Cd]}.
Since each operator $\hat a_k^\dagger$ and $\hat{d}^\dagger$ are linearly independent, we can split the above equation into two equalities as follows
\begin{eqnarray}\label{E_I}
\left( \epsilon_m-\varepsilon_k \right) u_{k,m} - \frac{W}{2L} \sum_q u_{q,m} = \frac{V}{\sqrt{2L}} u_{d,m},
\end{eqnarray}
and
\begin{eqnarray}\label{E_II}
 \left( \epsilon_m-\varepsilon_d  \right) u_{d,m} = \frac{V}{\sqrt{2L}}  \sum_q u_{q,m}.
\end{eqnarray}

To solve the system of equations defined by \eqref{E_I} and \eqref{E_II}, we can first isolate the term $u_{k,m}$ in \eqref{E_I} and apply a sum over all possible values of $k$, resulting in
\begin{eqnarray}\label{aux1}
 \left[1 - \frac{W}{2L} \sum_k \frac{1}{\epsilon_m - \varepsilon_k}\right] \sum_q u_{q,m} = \frac{V}{\sqrt{2L}} \sum_k \frac{1}{\epsilon_m - \varepsilon_k} u_{d,m}.  
\end{eqnarray}
Now, isolating the term $\sum_q u_{q,m}$ in equation \eqref{E_II} and substituting it into the equation \eqref{aux1} to find the following relation:
\begin{eqnarray}\label{Neu_Aux1}
 (\epsilon_m - \varepsilon_d) \left[1 - \frac{W}{2L} \sum_k \frac{1}{\epsilon_m - \varepsilon_k}\right]  = \frac{|V|^2}{2L} \sum_k \frac{1}{\epsilon_m - \varepsilon_k}.  
\end{eqnarray}

We already know that the sum $S_m = \dfrac{1}{2L} \mathlarger\sum_k \dfrac{1}{\epsilon_m - \varepsilon_k}$ can be computed by equation \eqref{soma}. Using this information in equation \eqref{Neu_Aux1}, the eigenvalues $\{\epsilon_m\}$ can implicitly be found through the expression
\begin{eqnarray}
 \epsilon_m = \varepsilon_m - \frac{\delta_m}{\pi}\Delta= \varepsilon_d + \frac{S_m |V|^2 }{1 - WS_m}, 
\end{eqnarray}
while $S_m $ depending only on $\delta_m$ and the energy $\varepsilon_m$ (that is, $S_m = S_m(\delta_m,\varepsilon_m)$). More specifically, there are two equalities in the above expression: 
\begin{equation}
    \varepsilon_m - \frac{\delta_m}{\pi}\Delta = \varepsilon_d + \frac{S_m|V|^2}{1 - WS_m},
\end{equation}
which  implicitly allows us to find the value of $\delta_m$ and $\epsilon_m = \varepsilon_m - \frac{\delta_m}{\pi}\Delta$, where we can find the eigenenergies $\{\epsilon_m\}$.

Considering $|\varepsilon_m - \varepsilon_d| \gg \frac{|\delta_m|}{\pi} \Delta$, after some mathematical manipulations, it is straightforward to show that the phase shift for the metallic levels can be obtained by
\begin{eqnarray}\label{phase_shift_new}
\tan \delta_m \approx \tan \delta_m^{(0)} { \left({1 +  \frac{\tan \delta_m^{(0)} }{\pi} \log\left(\frac{D+\varepsilon_l}{D  - \varepsilon_l}\right) }\right)^{-1} },
\end{eqnarray}
where 
\begin{eqnarray}
\tan \delta_m^{(0)}  = -\left[\pi\rho W  + \frac{\pi \rho V^2}{\varepsilon_m-\varepsilon_d} \right].
\end{eqnarray}

To complete the diagonalization procedure, we need to determine the coefficients $u_{k,m}$ and $u_{d,m}$. The basis transformation must satisfy the normalization condition, $\sum_k |u_{k,m}|^2 + |u_{d,m}|^2 = 1$. To apply this condition, it is convenient to first express $u_{k,m}$ in terms of $u_{d,m}$. One possible approach is to first isolate the term $\sum_q u_{q,m}$ in equation ~\eqref{aux1} resulting in
\begin{eqnarray}
  \sum_q u_{q,m} = \frac{V}{\sqrt {2L}} \left[1 - \frac{W}{2L} \sum_k \frac{1}{\epsilon_m - \varepsilon_k}\right]^{-1}\sum_k \frac{1}{\epsilon_m - \varepsilon_k} u_{d,m}.
\end{eqnarray}
Substituting the above relation into equation \eqref{E_I} and performing some algebraic manipulations, while keeping in mind the definition of $S_m$, we can rewrite the equation in the following form:  
\begin{eqnarray}\label{aux3}
u_{k,m} = \left[ \frac{V}{\sqrt {2L}} \frac{1}{\epsilon_m - \varepsilon_k}\right] \frac{1}{1 - WS_m} u_{d,m}.  
\end{eqnarray}

Finally, after expressing the coefficients $u_{k,m}$ in terms of $u_{d,m}$ as shown in equation \eqref{aux3},  we can square it and apply the summation over all possible $\{k\}$, resulting in the following expression
\begin{eqnarray}
\sum_k |u_{k,m}|^2 = \left[ |V|^2 \frac{1}{2L} \sum_k \frac{1}{(\epsilon_m - \varepsilon_k)^2} \right] \frac{1}{(1 - WS_m)^2} |u_{d,m}|^2.
\end{eqnarray}
The last step to conclude the Hamiltonian's diagonalization is to substitute the above expression into the normalization condition $\sum_k |u_{k,m}|^2 + |u_{d,m}|^2 = 1$, leading to 
\begin{eqnarray}\label{aux2}
\left[ \frac{1}{2L \Delta^2 }\left|\frac{\pi}{\sin \delta_m}\right|^2 \frac{|V|^2}{(1 - WS_m)^2} + 1\right] |u_{d,m}|^2 = 1.  
\end{eqnarray}
Here, we used the result in equation \eqref{Derivative_sum}. 

From the expression in equation \eqref{aux2}, we find the coefficients $u_{d,m}$, and consequently, we obtain $u_{k,m}$ using the equation \eqref{aux3}. Additionally, we can compute the eigenenergies $\epsilon_m$ by equation \eqref{Neu_Aux1} and an expression for the phase shift $\delta_m$ using equation \eqref{phase_shift_new}, which concludes the procedure. However, let us notice that $2 L \Delta = 2D =  1/\rho$, and taking $\Delta \rightarrow 0$, we can find that $\dfrac{\rho |V|^2 }{\Delta (1 - WS_m)^2} \left|\dfrac{\pi}{\sin \delta_m}\right|^2  \gg 1 $ for finite $V$. Therefore, we can write
\begin{eqnarray}
 |u_{d,m}| \approx \left[ \frac{1}{\sqrt{2L} \Delta }\left|\frac{\pi}{\sin \delta_m}\right| \frac{|V|}{(1 - WS_m)}\right]^{-1},
\end{eqnarray}
and a more friendly look form for the coefficients $u_{k,m}$
\begin{eqnarray}
u_{k,m} \approx -\frac{\Delta}{\left( \epsilon_m-\varepsilon_k \right)} \frac{\sin \delta_m}{\pi}.
\end{eqnarray}

In the case where $W=0$, we recover the pure non-interacting SIAM solution and equation \eqref{phase_shift_new} simplifies to:  
\begin{eqnarray}\label{phase_shift_imp}
\tan \delta_m \approx \frac{\Gamma}{\varepsilon_d - \varepsilon_m}{ \left({1+\frac{\Gamma}{\pi(\varepsilon_d - \varepsilon_k)} \log\left(\frac{D+\varepsilon_m}{D - \varepsilon_m}\right) }\right)^{-1} },
\end{eqnarray}
where $\Gamma = \pi \rho |V|^2$ is called hybridization function and quantifies how strongly the impurity is coupled to the metallic band.

\section{Analytical vs. numerical results}

After this point, we consider $\varepsilon_k = \varepsilon$ to simplify the notation, which is more appropriate for dealing with metallic bands as the energy levels tends to the continuum when $\Delta \rightarrow 0$. The phase shift of each final level of the metallic band can also be determined through “reverse engineering”, using the values of the final eigenenergies and the expression
\begin{equation}\label{phase_shif_num}
    \frac{\delta(\varepsilon)}{\pi} \equiv \frac{\varepsilon-\epsilon}{\Delta}.
\end{equation}
We use this expression to find numerically the phase shift. The energy $\epsilon$ can be identified in the numerical results by the corresponding final state that has the highest projection onto the initial state corresponding to the energy level $\varepsilon$.

Let us compare the analytical results obtained via the Sommerfeld-Watson transformation with the numerical results calculated through brute-force diagonalization of the discussed Hamiltonians, considering $ L = 2500 $. This procedure consists of numerically constructing the reference Hamiltonian and using numerical tools to diagonalize it.

\subsection{Fermi gas}

Usually, in calculations using the analytical solution presented here, the phase shift of the entire band is approximated by the phase shift near the Fermi energy ($\delta_0$). In this subsection, we verify the validity of the constant phase shift approximation in equation \eqref{phase_shift_ap} by comparing it with the phase shift obtained from equation \eqref{phase_shif_num} based on numerical results, and with the analytical phase shift obtained by equation \eqref{phase_shift}. Additionally, we compare the diagonal coefficients obtained from the constant phase shift approximation with those obtained through brute force diagonalization.

\begin{figure}[hbt!]
	\centering
        \includegraphics[scale=0.502]{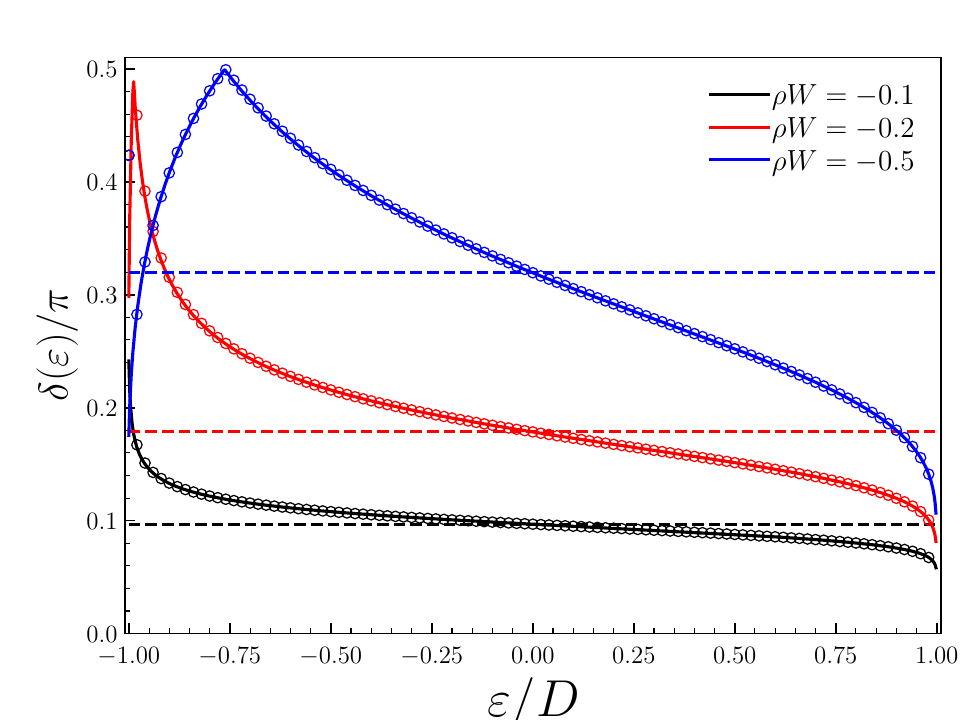}
        \includegraphics[scale=0.502]{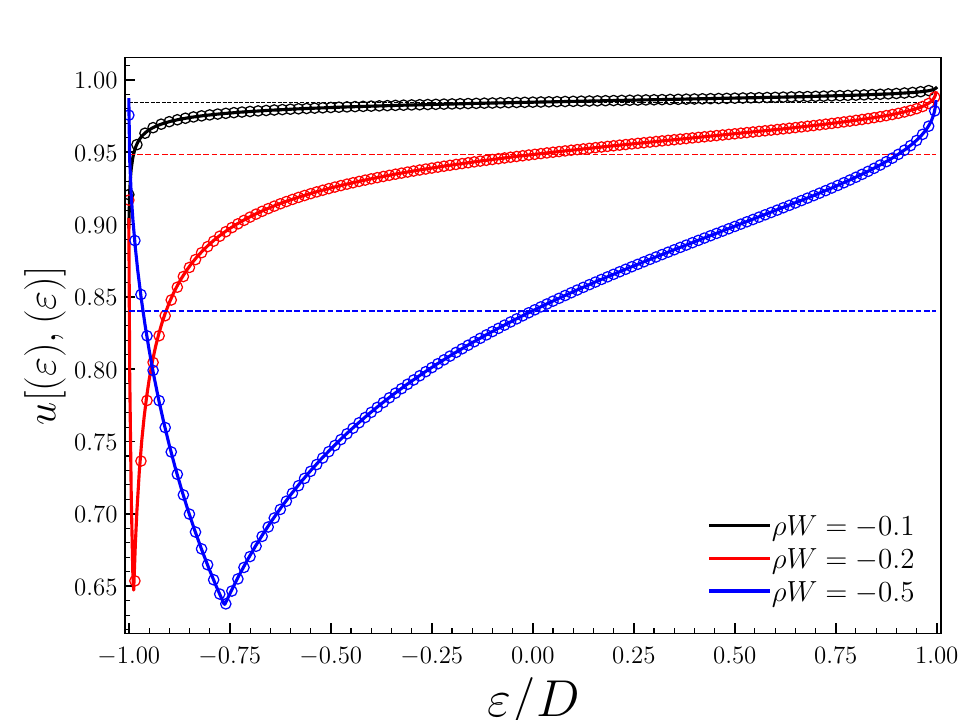}
        \vspace{-1.00cm}
	\caption{(Left panel) We show the values of $\delta(\varepsilon)$ computed by equation \eqref{phase_shift} (solid line), and by equation \eqref{phase_shift_ap} (dashed line), for different values of $\rho W$. (Right panel) We show the values of the diagonal coefficients computed by equation \eqref{coeficients} considering the $\delta(\varepsilon)$ in equation \eqref{phase_shift} (solid line), the values computed considering the $\delta$ constant (dashed line), and the numerical values computed by direct diagonalization (circular dots).}
        \label{Analytical_Coefficients}
\end{figure}

In Fig. \ref{Analytical_Coefficients}, the left panel displays the phase shift values. The solid line represents those calculated using equation \eqref{phase_shift_ap}, the circular dots correspond to the numerical results obtained from equation \eqref{phase_shif_num}, while the dashed line shows the phase shift values computed from equation \eqref{phase_shift} for various $\rho W$ values (black for $-0.1$, red for $-0.2$, and blue for $-0.5$). In the right panel, we show the values of the diagonal coefficients: the solid line represents those computed using equation \eqref{coeficients} with the correct phase shift, the dashed line shows values computed assuming a constant phase shift, and the circular dots represent the numerical values obtained through direct diagonalization. We observe a strong agreement between the expression derived from equation \eqref{coeficients} and the numerical results. Notably, for $\rho |W| \leq 0.1$, the phase shift constant approximation holds well up to high energies ($|\varepsilon| \le 0.25 D$). 

However, as observed in Fig. \ref{Analytical_Coefficients}, the deviations from the dashed line (constant phase shift approximation) become more significant when $\rho |W| \geq 0.2$. In these cases, assuming a constant phase shift is a poor approximation, and one must pay close attention to these details when working with such problems. Nevertheless, when considering long-time scales or low-energy regimes, where only the levels near the Fermi energy are relevant, the constant phase shift approximation remains a reasonable and effective simplification for analytical calculations.

Additionally, as $\rho W$ increases, distortion of the phase shift near the bottom of the band becomes evident in Fig. \ref{Analytical_Coefficients}. The origin of this distortion comes from the presence of the pole $\left({1 -  \rho W \log\left(\frac{D+\varepsilon}{D  - \varepsilon}\right) }\right)^{-1} $ of equation \eqref{phase_shift}, localized at the energy 
\begin{equation}
    \bar\varepsilon = D \cdot \mathrm{tanh} \left( \frac{1}{2\rho W}\right).
\end{equation}
For small values of $\rho |W|$, the singularity position of the $\tan \delta(\varepsilon)$ is $|\bar\varepsilon| \approx D$ (at one of the edges of the band), but for bigger values of $\rho |W|$ the singularity approaches the Fermi energy, as shown by the blue curve in Fig. \ref{Analytical_Coefficients}. Physically, this means that fermions in this energy region strongly feel the presence of the scattering potential and $|\delta(\bar\varepsilon)| \rightarrow \pi/2$.

We used negative values for the scattering potential to compare with the analytical results. However, it can be easily observed that changing the sign of $ W $ only mirrors the results with respect to both the $ x $- and $ y $-axes. This symmetry arises from the invariance under the transformations  $\delta \rightarrow -\delta $, $ W \rightarrow -W $ and $ \varepsilon \rightarrow -\varepsilon $, which can be verified by equation \eqref{phase_shift}.

\subsection{Non-interacting SIAM}

\begin{figure}[hbt!]
	\centering
         \includegraphics[scale=0.50]{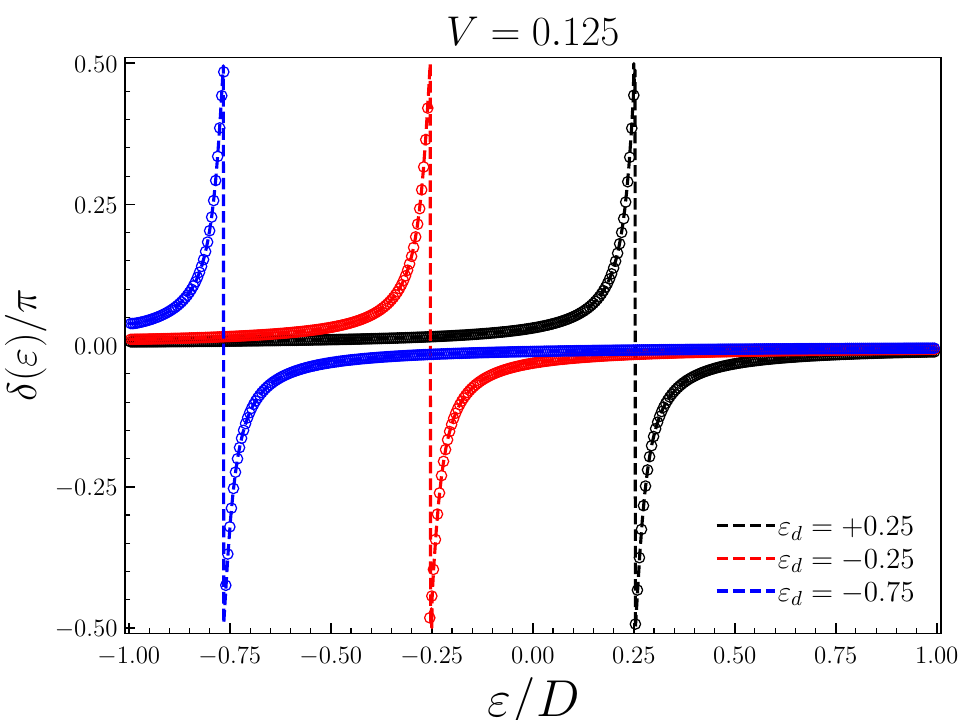}
        \includegraphics[scale=0.50]{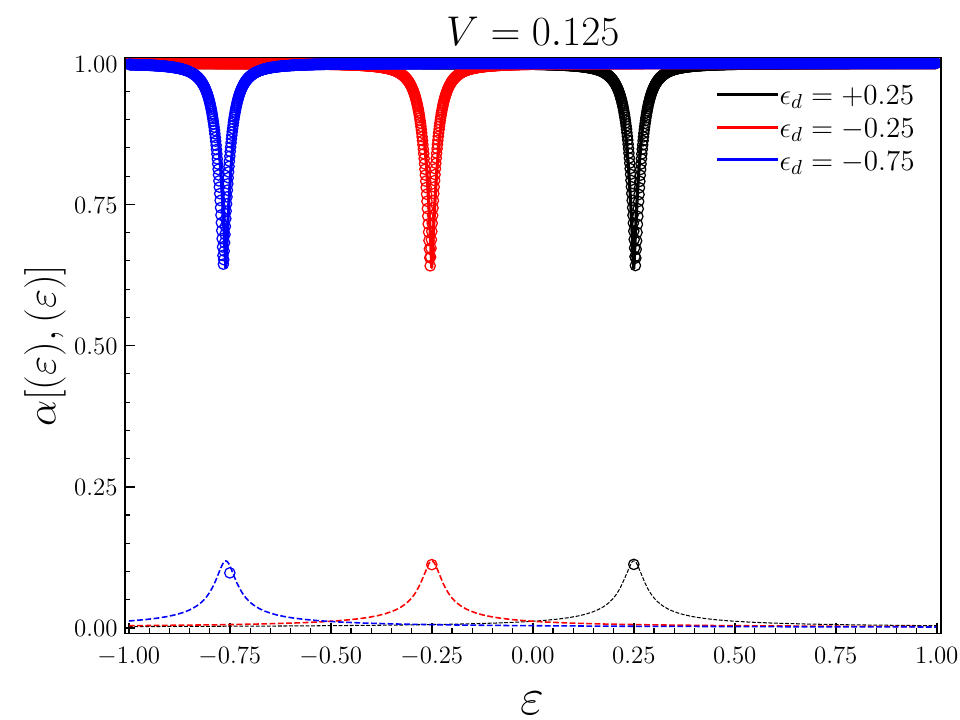}
        \includegraphics[scale=0.50]{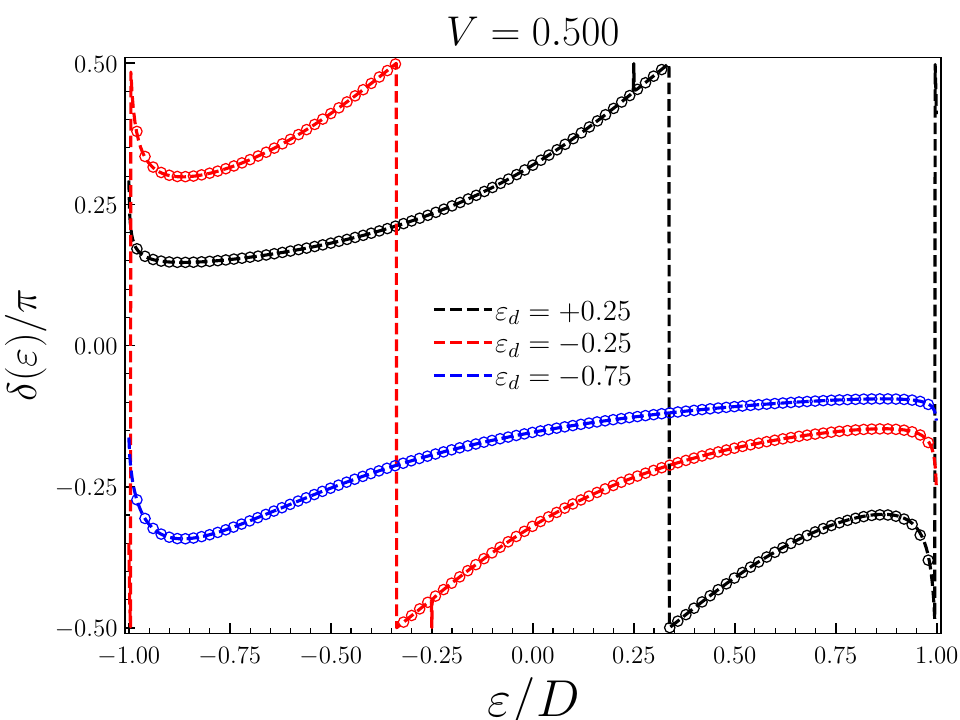}
        \includegraphics[scale=0.50]{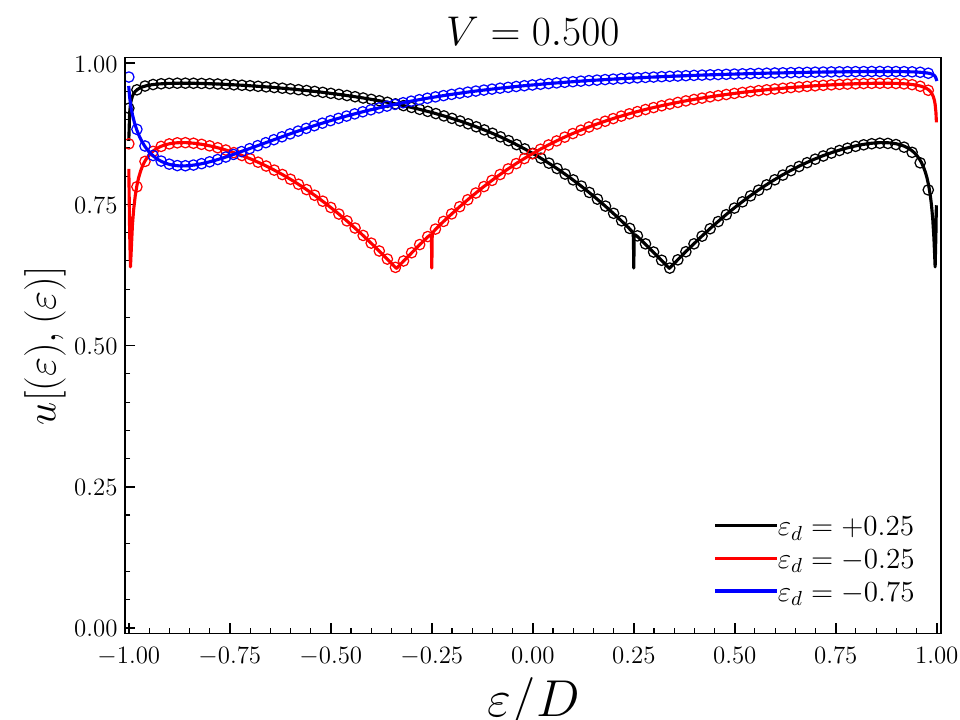}
        \vspace{-1.1cm}
	\caption{(Top panels) Phase shift for different values of $\varepsilon_d$ and $V$ (values shown in the plots).  (Bottom panels) Diagonal coefficients for different values of $\epsilon_d$ and $V$ computed computationally (circular dots) and by analytical equations above (dashed lines). All energies shown in this plot are in unities of $D$. Clearly, the phase shift is stronger close to the impurity level, and the decay ratio depends on $V$ (or $\Gamma$). }
        \label{Phase_Shift_Coeff_Impurity}
\end{figure}

Now, let us discuss the disturbances in the metallic band caused by the presence of a single impurity and compare the analytical results (from equations \eqref{phase_shift_imp} and \eqref{Analytical_Coefficients}) with those obtained through brute force diagonalization. In Fig. \ref{Phase_Shift_Coeff_Impurity}, we show the results for the phase shift (left panels) and the diagonal coefficients (right panels) for two different values of $ V $: $ V = 0.125 D $ (top panels) and $ V = 0.500 D $ (bottom panels). The calculations are performed for $\varepsilon_d = -0.75D$ ({blue~curves}),  $\varepsilon_d = -0.25D $  ({red~curves}), and $\varepsilon_d = +0.25D$ ({black~curves}). Once again, the analytical results worked extremely well, except for very near the singularities, where small discrepancies can be noticed. Another interesting observation is that the phase shift becomes more pronounced near the impurity level, approaching $\pm \pi/2$, and its decay rate is influenced by $V$ (or $\Gamma$).

As expected for impurity-metal systems, the band fermionic levels near the impurity energy level are strongly perturbed by the presence of the impurity. For $ V = 0.125 D$ (or $ \Gamma = 0.025 D $), the results are similar for any value of $ \varepsilon_d $: a sharp peak appears near $ \varepsilon_d $, where the phase shift of the nearby electronic states tends to $ \pm \frac{\pi}{2} $, followed by a rapid decay as the energy moves away from $ \varepsilon_d $. This sharp peak results from the pole $ \left[\varepsilon_d - \varepsilon + \frac{1}{\pi}\Gamma  \log\left(\frac{D+\varepsilon}{D  - \varepsilon}\right) \right]^{-1} $ in equation \eqref{phase_shift_imp}. If $\Gamma$ is small, the position of the singularity, 
\begin{equation}
    \bar\varepsilon \approx \varepsilon_d + \frac{1}{\pi} \Gamma \log\left(\frac{D+\varepsilon_d}{D  - \varepsilon_d}\right),
\end{equation} 
is close to the impurity energy level ($\bar \varepsilon \approx \varepsilon_d$). However, for larger values of $ \Gamma $, the peak position is significantly shifted, as shown in Fig. \ref{Phase_Shift_Coeff_Impurity} for $ V = 0.500D $ (or $ \Gamma = 0.393D $). Additionally, the decay becomes slower, indicating that hybridization effects extend farther from the impurity energy level. The same discussion regarding the phase shift also applies directly to the diagonal projections, which exhibit a similar behavior. However, instead of a singularity, the diagonal projections show a minimum at the corresponding point.

\section{A more general result for the NI-SIAM}

Note that all examples discussed here (see equations \eqref{phase_shift}, \eqref{phase_shift_new} and \eqref{phase_shift_imp}) result in the unified analytical solution for the phase shift and coefficient as  
\begin{eqnarray}
\tan \delta_m = \tan \delta_m^{(0)} { \left({1 +  \frac{\tan\delta_m^{(0)}}{\pi} \log\left(\frac{D+\varepsilon}{D  - \varepsilon}\right) }\right)^{-1} }
\end{eqnarray}
and 
\begin{eqnarray}\label{general:coefficients}
u_{k,m} = -\frac{\Delta}{\left( \varepsilon_m -\varepsilon_k  - \frac{\delta_m}{\pi}  \Delta \right)} \frac{\sin \delta_m}{\pi}.
\end{eqnarray}
Here, the differences in the Hamiltonians lie solely in the general phase shift, particularly, in the term $\tan \delta_m^{(0)}$. In fact, The term  $\tan \delta_m^{(0)}$ can be summarized in this unified solution as
\begin{eqnarray}
&\tan \delta_m^{(0)} & = -\pi \rho W, ~~~~~~~~~~~~~~~~~~(\mathrm{Scattering~potential}); \nonumber \\
&\tan \delta_m^{(0)}& = \frac{\Gamma}{\varepsilon_d - \varepsilon_m}, ~~~~~~~~~~~~~~~~(\mathrm{Impurity});  \nonumber \\
&\tan \delta_m^{(0)}& = -\pi \rho W + \frac{\Gamma}{\varepsilon_d - \varepsilon_m},  ~~~(\mathrm{Impurity~plus~scattering~potential});  
\end{eqnarray}
which quantifies the strength of the disturbances experienced by the metallic band due to the presence of the impurity and/or an external scattering potential. 

Motivated by this simplified solution, we write down a more generic Hamiltonian as 
\begin{equation}\label{General:HMain}
   \hat H =\varepsilon_d \hat d^\dagger \hat d +\sum_{k}\varepsilon_{k} \hat a^\dagger_{k}\hat a_k + \frac{V}{\sqrt{2L}} \left( \hat d^\dagger \sum_{k} \alpha_{k} \hat a_k  +{\mathrm H.c.} \right) + \frac{W}{2L} \sum_{k} \alpha_k^* \hat a_k^\dagger \sum_q \alpha_q \hat a_q.
\end{equation}
Here, $\alpha_k$ satisfies the normalization condition $\sum_k |\alpha_k|^2 = 2L$, where $2L$ is the total number of energy levels. The transformation $\alpha_k \rightarrow 1$ allows us to recover the Hamiltonians previously discussed. The dispersion $\varepsilon_k$ is now generic, and the density of states can be computed generically as:  
\begin{equation}\label{General:DOS}
   \rho(\varepsilon) = \frac{1}{2L} \left[ \dv{\varepsilon}{k}  \right]^{-1} .
\end{equation}
This definition may seem slightly different from the usual one, but the inclusion of the term $\frac{1}{2L}$ ensures that the density of states remains independent of the system size.

Using this more generic Hamiltonian, and the general density of states in equation \eqref{General:DOS}, Appendix \ref{General_Solution} shows that the phase shift can be written in a similar form as
\begin{align}\label{General:phaseshift}
    \tan \delta_m = \tan \delta^{(0)}_m \left( 1 + \frac{\tan \delta^{(0)}_m}{\pi } \frac{1}{\rho(\varepsilon_m)} \mathcal{P}\int_{-D}^{+D} {d\epsilon} ~\frac{ \rho(\varepsilon) |\alpha_\varepsilon|^2 }{\varepsilon_m-\varepsilon } \right)^{-1}. 
\end{align}
where
\begin{align}\label{General:phaseshift_}
    \tan \delta^{(0)}_m =  - \pi W \rho(\varepsilon_m)  ~  |\alpha_m|^2 + \frac{\Gamma(\varepsilon_m) }{\varepsilon_m-\varepsilon_d}, 
\end{align}
with $\Gamma(\varepsilon) = \pi \rho(\varepsilon) V^2$. One can easily verify that, by considering $\varepsilon_k = \Delta \cdot k$ and $\alpha_\varepsilon = \alpha_k = 1$, the above equation simplifies to the already-present familiar form. As demonstrated in Appendix \ref{General_Solution}, the coefficients can also be approximated using equation \eqref{general:coefficients}, with $\Delta \rightarrow \Delta_m$, where
\begin{equation}
   \Delta_m =  \frac{1}{{2L\rho(\varepsilon_m)}} \frac{1}{|\alpha_m|}.
\end{equation}
This completes the diagonalization procedure for this more general Hamiltonian.

Finally, the last generalization concerns the dimensionality of $\vec{k}$-space. As previously mentioned, our focus here is on 1D systems. However, the results can be easily extended to higher dimensions that exhibit spherical symmetry in $d$ dimensions. In this case, the energy depends only on $k = |\vec{k}|$, and the summations $\sum_{\vec{k}}$ transform into sums $\sum_k \mathcal{G}_k$ over $k$, where $\mathcal{G}_k$ accounts for the total number of $\vec{k}$ points with magnitude $k$. These contributions are encapsulated in the density of states for $d$ dimensions, requiring only the replacement $(2L) \rightarrow (2L)^d$ in the equations presented in this section and the computation of the corresponding density of states $\rho\left(\varepsilon_{\left|\vec{k}\right|}\right)$.

\section{Conclusions}

In this work, we demonstrated the application of the Sommerfeld-Watson transformation in the diagonalization procedure of two fundamental and significant Hamiltonians in condensed matter physics: the Fermi gas and the non-interacting single impurity Anderson model. These models, despite their simplicity, exhibit rich physical phenomena. The calculations presented here were already performed by the authors cited in the introduction. In this paper, we provide an expanded explanation of the procedure, utilizing a more modern notation that is not readily available in the existing literature. We also compared the analytical results and the numerical results from brute-force diagonalization of these models for different parameters, which reveals excellent concordance between these results.

We also considered a more general non-interacting single-impurity Anderson model, where we did not assume a specific dispersion for the band energies and allowed the couplings in the Hamiltonian to vary depending on the band levels they couple to. The same procedure used for the Fermi gas and the simpler NI-SIAM has been shown to successfully diagonalize this more complex model, despite the increased mathematical complexity. To the best of our knowledge, no previous work has derived the expression we obtained for this more general model. Furthermore, we believe that the same procedure can be extended to higher-dimensional problems.

Lastly, we want to emphasize that, despite the simplicity of these models, they have historically driven the discovery of important new phenomena, as already discussed in the text. In modern times, while analytical calculations rarely lead to entirely new physics, complex Hamiltonians can often be simplified to cases with known analytical solutions. These simplifications can provide valuable insights into the underlying more complex phenomena.

\section*{Acknowledgments}

The authors would like to thank Prof. Dr. L. N. Oliveira for introducing and explaining this method to us for the first time. The authors would also like to acknowledge Prof. Dr. L. N. Oliveira for his supervision and for all the knowledge he has shared with us. G. Diniz acknowledges a PhD scholarship from the Brazilian agency Coordenação de Aperfeiçoamento de Pessoal de Nível Superior (CAPES - grant No. 88887.495890/2020-00). F. D. Picoli acknowledges PhD and internship fellowships from Fundação de Amparo à Pesquisa do Estado de São Paulo (FAPESP - grants 2022/09312-4 and 2024/05637-1). M. P. Lenzarini acknowledges funding from CAPES (Grant No. 88887.941429/2024-00).

\bibliography{main}

\newpage
\appendix

\section{Finding the commutators}\label{Commutator,General}

Here, we aim to compute a general commutator that can be used to derive the main commutators presented in the text. In our notation, $[\hat{A}, \hat{B}]$ denotes the commutator, while $\{\hat{A}, \hat{B}\}$ represents the anti-commutator. Since we are dealing with fermionic operators $\hat{c}_k^\dagger$, they obey the relation $\{\hat{c}_q, \hat{c}_k^\dagger\} = \delta^K_{q,k}$, where $\delta^K_{q,k}$ is the Kronecker delta. This general commutator can be obtained following the above calculation:
\begin{align}\label{Commut}
    \left[ \sum_{q,q'} M_{q,q'} \hat c_{q}^\dagger \hat c_{q'} ,   \hat c_k^\dagger \right] &=& &\sum_{q,q'} M_{q,q'}  \left[ \hat c_{q}^\dagger \hat c_{q'} ,   \hat c_k^\dagger \right] \nonumber, \\
    &=& &\sum_{q,q'} M_{q,q'}  \left[ \hat c_{q}^\dagger \hat c_{q'} \hat c_k^\dagger - \hat c_k^\dagger    \hat c_{q}^\dagger \hat c_{q'}  \right] \nonumber, \\
    &=& &\sum_{q,q'} M_{q,q'}  \left[ \hat c_{q}^\dagger \left( \delta^K_{k,q'} -  \hat c_k^\dagger \hat c_{q'} \right) - \hat c_k^\dagger    \hat c_{q}^\dagger \hat c_{q'}  \right] \nonumber,  \hspace{3cm}\\
    &=& &\sum_{q,q'} M_{q,q'}  \left[ \hat c_{q}^\dagger  \delta^K_{k,q'} -  \hat c_{q}^\dagger  \hat c_k^\dagger \hat c_{q'} - \hat c_k^\dagger    \hat c_{q}^\dagger \hat c_{q'}  \right]  \nonumber, \\
    &=& &\sum_{q,q'} M_{q,q'}  \left[ \hat c_{q}^\dagger  \delta^K_{k,q'} - \left\{ \hat c_{q}^\dagger ,  \hat c_k^\dagger \right\}  \hat c_{q'} \right] \nonumber, \\
    &=& &\sum_{q} M_{q,k}  \hat c_{q}^\dagger.
\end{align}
This result in equation \eqref{Commut} can be used to easily derive all commutators used in this work, one only needs to choose the coefficient $M_{q,q'}$ accordingly.

\section{Finding the eigenvalue and eigenoperators}\label{Eigenvalues_and_Eigenvectors}

Using the equality defined in equation \eqref{[H,gm]_test} by substituting $\hat g_m^\dagger$ with the linear combination defined in equation \eqref{gm_test} and after some algebraic manipulation, we find
\begin{eqnarray}
\sum_k \left[\varepsilon_k u_{k,m} +  M_{k} \sum_q \bar M_{q} u_{q,m}  +   \bar M_{k}^* \sum_q  M_{q}^* u_{q,m}  \right] \hat c_k^\dagger = \sum_k \epsilon_m u_{k,m} \hat c_k^\dagger.
\end{eqnarray}
As the operators $ \hat c_k^\dagger$ are linearly independent, each term of the $k$-sum above has to satisfy the above equation independently,
\begin{eqnarray}\label{E__I_test_ap}
 u_{k,m}  = \frac{M_k}{\left( \epsilon_m-\varepsilon_k \right)}  \sum_q \bar M_{q} u_{q,m}  +  \frac{\bar M_k^*}{\left( \epsilon_m-\varepsilon_k \right)}  \sum_q  M_{q}^* u_{q,m} .
\end{eqnarray}
After multiplying both sides of equation \eqref{E__I_test_ap} by $\bar M_{k}$ summing over all $k$ states, we find
\begin{eqnarray}\label{Conection_ap}
\left[ 1 -  \sum_k \frac{ \bar M_k M_k}{\epsilon_m-\varepsilon_k} \right] \sum_q \bar M_{q}  u_{q,m}  =\sum_k \frac{| \bar M_k|^2}{\epsilon_m-\varepsilon_k} \sum_k  M_{q}^* u_{q,m},
\end{eqnarray}
and now multiplying both sides of equation \eqref{E__I_test_ap} by $M_{k}^*$ summing over all $k$ states, we find
\begin{eqnarray}\label{Conection_ap_}
\left[ 1 -  \sum_k \frac{ \bar M_k^* M_k^*}{\epsilon_m-\varepsilon_k} \right] \sum_q M_{q}^* u_{q,m}  =\sum_k \frac{| M_k|^2}{\epsilon_m-\varepsilon_k} \sum_k  \bar M_{q}  u_{q,m}.
\end{eqnarray}

Considering that the terms $|\sum_q M_{q}^* u_{q,m}|$ and $|\sum_q \bar{M}_{q} u_{q,m}|$ are nonzero, the system of equations defined in equations \eqref{Conection_ap} and \eqref{Conection_ap_} leads to
\begin{eqnarray}\label{Conection*_ap}
\left| 1 -  \sum_k \frac{ \bar M_k M_k}{\epsilon_m-\varepsilon_k} \right|^2  = \sum_k \frac{| M_k|^2}{\epsilon_m-\varepsilon_k} \sum_k \frac{| \bar M_k|^2}{\epsilon_m-\varepsilon_k}.
\end{eqnarray}

The equation defined in equation \eqref{Conection*_ap} represents the eigenvalue equation, and the set of values $ \{\epsilon_m\} $ that satisfy this equation are the eigenvalues of the Hamiltonian. This is precisely where the Sommerfeld-Watson transformation becomes highly useful, as it allows us to find functions that simplify the summations 
\begin{equation}\label{Motiv:_ap}
    S(\epsilon) = \sum_{k} \dfrac{|M_k|^2}{\epsilon-\varepsilon_k},~~~~ \bar S(\epsilon) = \sum_{k} \dfrac{|\bar M_k|^2}{\epsilon-\varepsilon_k}, ~~ \mathrm{and}~~  \tilde S(\epsilon) = \sum_{k} \dfrac{M_k \bar M_k}{\epsilon-\varepsilon_k},
\end{equation}
and the solutions to the equation $\left| 1- \tilde S(\epsilon)\right|^2 = S(\epsilon)\tilde S(\epsilon)  $ give to us the eigenvalues of $\hat H$.

Manipulating equation \eqref{E__I_test_ap} and imposing the normalization condition $ \sum_k |u_{k,m}|^2 = 1 $, it is straightforward to show that 
\begin{equation}\label{Motiv_ap}
    1 = - \dv{S(\epsilon)}{\epsilon}\left|\sum_q \bar M_q u_{q,m}\right| ^2 -\dv{\bar{S}(\epsilon)}{\epsilon} \left|\sum_q M_q^* u_{q,m}\right| ^2  -2\mathrm{Re} \left(\dv{\tilde{S}(\epsilon)}{\epsilon} \sum_q M_q u_{q,m} \sum_q \bar M_q u_{q,m}\right).
\end{equation}
Finding these summations over $k$ enables us to determine the terms $ \sum_q \bar{M}_q u_{q,m} $ and  $ \sum_q {M}_q^* u_{q,m} $, which in turn allows us to calculate $ u_{q,m} $, thereby completing the diagonalization procedure.

\section{A more general solution}\label{General_Solution}

We can write the Hamiltonian in equation \eqref{General:HMain} compactly as 
\begin{equation}\label{General:H}
    \hat H = \sum_{\kappa} \varepsilon_\kappa \hat{c}_\kappa^\dagger \hat{c}_\kappa + \sum_{\kappa,\xi}\left( M_{\kappa} \bar M_{\xi} +  M_\xi^* \bar M_\kappa^*  \right) \hat{c}_\kappa^\dagger \hat{c}_{\xi}.
\end{equation}
Here, the index $\kappa$ runs over the impurity level $d$ and band levels $k$, the index $\xi$ runs over $d$ and band levels $q$, and the coupling can be expressed as
\begin{equation}\label{General:Coupling}
    M_{\kappa} = \sqrt\frac{{W}}{{4L}} (1- \delta^K_{\kappa,d}) \alpha_k^* ~\mathrm{and} ~ \bar M_{\xi} = \frac{V}{\sqrt{W/2}} \delta^K_{\xi,d} + \sqrt\frac{{W}}{{4L}}  (1- \delta^K_{\xi,d}) \alpha_q, 
\end{equation}
where $\delta^K_{\kappa,\xi}$ stands for the Kronecker's delta.

\subsection{The general phase shift}

The compact Hamiltonian in equation \eqref{General:H} allows us to directly use all equations deduced in the previous appendix, particularly the eigenvalue equation in equation \eqref{Conection*}. Using the couplings defined by equation \eqref{General:Coupling} leads to:
\begin{eqnarray}\label{AUX:G:1}
    S(\epsilon) & = & \sum_{k} \dfrac{|M_k|^2}{\epsilon-\varepsilon_k} = \frac{W}{4L} \sum_{k} \frac{|\alpha_k|^2}{\epsilon-\varepsilon_k}, \nonumber \\
    \bar{S}(\epsilon) & = & \sum_{k} \dfrac{|\bar M_k|^2}{\epsilon-\varepsilon_k} = \frac{2 V^2/ W}{\epsilon-\varepsilon_d} + S(\epsilon), \nonumber \\
    \tilde{S}(\epsilon) & = & \sum_{k} \dfrac{M_k \bar M_k}{\epsilon-\varepsilon_k} =  S(\epsilon)
\end{eqnarray}
and
\begin{equation}\label{General:1}
    \left| 1 - \frac{W}{4L}\sum_{k} \frac{|\alpha_k|^2}{\epsilon-\varepsilon_k} \right|^2 = \left[ \frac{2 V^2/ W}{\epsilon-\varepsilon_d}  + \frac{W}{4L} \sum_{k} \frac{|\alpha_k|^2}{\epsilon-\varepsilon_k}  \right] \frac{W}{4L} \sum_{k} \frac{|\alpha_k|^2}{\epsilon-\varepsilon_k}.
\end{equation} 
A more friendly notation is obtained by substituting $|\alpha_k|^2$ by $w_k$.

We observe that the term $ S_0(\epsilon) = \frac{1}{2L} \sum_{k} \frac{w_k}{\epsilon - \varepsilon_k} $ appears repeatedly in the equations above. Thus, our solution relies on evaluating this summation. To procedure after this point, some assumptions are necessary. Fist, we will assume that $w_{k}$ can be expressed as a function $w(k)$ and it do not have poles. The Sommerfeld-Watson transformation then allowed us to express the summation as
\begin{align}\label{General:2}
    S_0(\epsilon) = \frac{1}{2L}\sum_{k} \frac{w_{k}}{\epsilon-\varepsilon_k} =  \frac{1}{2i} \oint \frac{dz}{2L} \frac{w(z) \cot(\pi z)}{\epsilon-\varepsilon_z } - \frac{\pi w(\epsilon) }{2L}  \cot(\pi z(\epsilon)) \mathrm{Res}\left(  \frac{1}{\epsilon-\varepsilon_z} \right)_{\varepsilon_z = \epsilon},
\end{align}
or
\begin{align}\label{General:3}
    S_0(\epsilon) = \frac{1}{2i} \oint \frac{dz}{2L} \frac{w(z) \cot(\pi z)}{\epsilon-\varepsilon_z } - \frac{\pi w(\epsilon)}{2L}  \cot(\pi z(\epsilon)) \mathrm{Res}\left(  \frac{1}{\epsilon-\varepsilon_z} \right)_{\varepsilon_z = \epsilon}.
\end{align}
Let us recall that the variable $z$ is a continuous complex extension of the discrete variable $k$, which labels the fermionic energy level of the band $\varepsilon_k$.

By using the same complex counter defined in Fig. \ref{Sommerfeld-Watson} and the density of states defined in equation \eqref{General:DOS} with no poles near the energy $\epsilon$, it is straightforward to show that
\begin{align}\label{General:4}
    S_0(\epsilon)  = \mathcal{P}\int_{-D}^{+D} {d\epsilon} ~\frac{ \rho(\varepsilon) w_\varepsilon }{\epsilon-\varepsilon } - \pi \rho(\epsilon) w(\epsilon) \cot(\pi z(\epsilon)) ~ \mathrm{Res}\left(  \frac{1}{2 L \rho(\epsilon) } \frac{1}{\epsilon-\varepsilon_z} \right)_{\varepsilon = \epsilon}.
\end{align}

A generic dispersion can be expressed as $\varepsilon_k = \gamma \cdot k^\beta$. Then, from equation \eqref{General:DOS}, the density of states is given by $\rho(\varepsilon_k) = \frac{1}{2L} \left(\gamma \beta k^{\beta-1} \right)^{-1}$. With this in mind, let us compute the following limit
\begin{align}
   \lim_{z \rightarrow z(\epsilon)} \frac{1}{2L\rho(\varepsilon_z)}  \frac{z-z(\epsilon)}{\varepsilon_z - \epsilon} = \lim_{z \rightarrow z(\epsilon)}  \frac{1}{2L\rho(\varepsilon_z)}  \frac{z-z(\epsilon)}{\gamma \cdot z^\beta - \epsilon} = \lim_{z \rightarrow z(\epsilon)}  \frac{1}{2L\rho(\varepsilon_z)} \left(\gamma \beta k^{\beta-1} \right)^{-1} = 1.
\end{align}
Clearly, the pole is simple around the energy $\epsilon$ and the residue can be computed by the above expression, resulting in
\begin{equation}
    \mathrm{Res}\left(  \frac{1}{2 L \rho(\epsilon) } \frac{1}{\epsilon-\varepsilon_z} \right)_{\varepsilon = \epsilon} = 1.
\end{equation}

Now, let as split the function $S_0(\epsilon) = \mathcal{S} (\epsilon) + \rho(\epsilon) \mathcal{I}(\epsilon)$ by defining 
\begin{align}\label{General:Integral}
    \mathcal{I}(\epsilon) = \frac{1}{\rho(\epsilon)} \mathcal{P}\int_{-D}^{+D} {d\epsilon} ~\frac{ \rho(\varepsilon) w_\varepsilon }{\epsilon-\varepsilon },
\end{align}
and 
\begin{align}\label{General:Approximation}
    \mathcal{S} (\epsilon)  = - \pi \rho(\epsilon) w(\epsilon) \cot(\pi z(\epsilon)).
\end{align}

Now, returning to equation \eqref{General:1}, we can write
\begin{align}\label{General:Aux}
    1 - W\left( \mathcal{S} (\epsilon) + \rho(\epsilon) \mathcal{I}(\epsilon) \right) = \frac{V^2}{\epsilon-\varepsilon} \left( \mathcal{S} (\epsilon) + \rho(\epsilon) \mathcal{I}(\epsilon) \right),
\end{align}
resulting in the expression 
\begin{align}\label{General:AuxI}
    \mathcal{S}(\epsilon)  \left(  \frac{V^2}{\epsilon-\varepsilon}  + W   \right) = 1 - \rho(\epsilon) \left(  \frac{V^2}{\epsilon-\varepsilon}  + W   \right) \mathcal{I}(\epsilon).
\end{align}

Finally, once the pole $\epsilon - \varepsilon_k$ results in a $z(\epsilon)$ near $z=k$, we can write $z(\epsilon) = k - {\delta(\epsilon)}/{\pi}$, and our expression for the phase shift becomes
\begin{align}\label{General:5}
    \tan \delta(\epsilon) = \tan \delta^{(0)}(\epsilon) \left( 1 + \frac{\tan \delta^{(0)}(\epsilon)}{\pi } \mathcal{I}(\epsilon) \right)^{-1}. 
\end{align}
where
\begin{align}\label{General:6}
    \tan \delta^{(0)}(\epsilon) =  - \pi \rho(\epsilon) W~ w(\epsilon) + \frac{\Gamma(\epsilon) }{\epsilon-\varepsilon_d}. 
\end{align}

\subsection{The general coefficients}

As shown in Appendix \ref{Eigenvalues_and_Eigenvectors}, the coefficients can be determined using the auxiliary equation \eqref{Motiv_ap}. As observed in equation \eqref{AUX:G:1}, all the key summations can be obtained by computing $S_0$. Consequently, to extract information from equation \eqref{Motiv_ap}, we first need to compute its derivative in the following expression
\begin{align}
    -\frac{1}{2L}\dv{S_0(\epsilon)}{\epsilon}  = \frac{\pi^2 \rho^2(\epsilon)}{  \sin^{2}(\delta(\epsilon))} w(\epsilon) ~ +\frac{\pi}{2L}\cot(\delta(\epsilon)) \dv{\epsilon}\left(\rho(\epsilon) w(\epsilon) \right) -\frac{1}{2L}\dv{\epsilon} \left(\mathcal{P}\int_{-D}^{+D} {d\epsilon} ~\frac{ \rho(\varepsilon) w_\varepsilon }{\epsilon-\varepsilon }\right).
\end{align}
Here, the fraction $1/2L$ was inserted to show how much larger the first term on the left side of the above equation is compared to the other terms. As a result, we can drop these lesser contributions, resulting in
\begin{align}\label{Derivative_S0}
    -\frac{1}{2L}\dv{S_0(\epsilon)}{\epsilon}  =& \pi^2 \rho^2(\epsilon)  \sin^{-2}(\delta(\epsilon)) w(\epsilon)  + \mathcal{O}\left((2L)^{-1}\right).
\end{align}

For simplicity, let us define $x_1 = \sum_q \bar M_{q}  u_{q,m}  $ and $x_2 = \sum_q M_{q}^*  u_{q,m} $, transforming the system of equations formed by equations \eqref{Conection_ap_} and \eqref{Motiv_ap} into
\begin{eqnarray}\label{Conection_ap_2}
\left[ 1 - S(\epsilon) \right] x_2  =  S(\epsilon)~ x_1 ,
\end{eqnarray}
and
\begin{equation}
    1 = - \dv{S(\epsilon)}{\epsilon}x_1^2 +\left[ \frac{2 V^2/ W}{(\epsilon-\varepsilon_d)^2} - \dv{S(\epsilon)}{\epsilon}  \right] x_2^2  -2\dv{S(\epsilon)}{\epsilon} x_1 x_2,
\end{equation}
respectively. After grouping the common terms $\dv{S}{\epsilon}$ of the above equation, we obtain
\begin{equation}\label{G:AUX:N}
    1 = - \dv{S(\epsilon)}{\epsilon} \left( x_1+ x_2 \right)^2 +\left[ \frac{2 V^2/ W}{(\epsilon-\varepsilon_d)^2} \right] x_2^2.
\end{equation}
Now, we only need a relation between $x_1$ and $x_2$ to finally find them.

Keeping that in mind, isolating $x_2$ from equation \eqref{Conection_ap_2} and substituting into equation \eqref{G:AUX:N} results in the following expression
\begin{equation}
    1 = \left[- \frac{d S(\epsilon)}{d\epsilon} + \frac{2 V^2/ W}{(\epsilon-\varepsilon_d)^2}  S^2(\epsilon) \right] (x_1 + x_2)^2.
\end{equation}

Now, by substituting $S(\epsilon) = \frac{W}{2} S_0$ and using the result found in equation \eqref{Derivative_S0}, it is straightforward to show that
\begin{equation}
    1 = \left[ \sqrt{\frac{2L W}{2}}\frac{\pi \rho(\epsilon)}{\sin (\delta(\epsilon)) } \sqrt{w(\epsilon) }  \right]^2 \left( 1 + \mathcal{O}\left((2L)^{-2}\right) \right)  (x_1 + x_2)^2,
\end{equation}
and consequently the expression
\begin{equation}\label{G:AUX:X1X2}
    (x_1 + x_2) \approx \left[ \sqrt{\frac{2LW}{2}}\frac{\pi \rho(\epsilon)}{\sin (\delta(\epsilon)) } \sqrt{w(\epsilon) }  \right]^{-1} .
\end{equation}

A combination of equation \eqref{E__I_test_ap} with the general couplings defined in equation \eqref{General:Coupling} results in the following expression
\begin{equation}\label{G:AUX:COE}
     u_{k,m}  = \sqrt{\frac{W}{4L}}\frac{(x_1+x_2)}{\left( \epsilon_m-\varepsilon_k \right)},
\end{equation}
which depends exactly on the quantity $x_1 + x_2$ already found.

Continuing the calculations, the general coefficients can be obtained by substituting equation \eqref{G:AUX:X1X2} into equation \eqref{G:AUX:COE}, resulting in
\begin{equation}
     u_{k,m}  \approx - \frac{1}{{2L}} \frac{1}{\sqrt{w(\epsilon) }}\frac{\sin (\delta(\epsilon_m))}{\pi \rho(\epsilon_m)} \frac{1}{\epsilon_m-\varepsilon_k}.
\end{equation}

Finally, by defining the quantity
\begin{equation}
     \Delta_m = \frac{1}{{2L\rho(\epsilon_m)}} \frac{1}{\sqrt{w(\epsilon) }},
\end{equation}
the expression for the coefficients can be written in the familiar form as
\begin{equation}
     u_{k,m}  \approx -\frac{\sin (\delta(\epsilon_m))}{\pi} \frac{\Delta_m}{\epsilon_m-\varepsilon_k}.
\end{equation}
Completing the diagonalization procedure for this more general Hamiltonian.

\end{document}